\documentclass[sigconf]{acmart}

\newcommand{\user}[1]{\textsf{\textit{``#1''}}}

\newcommand{\agent}[1]{%
    \begingroup
    \spaceskip=0.2em plus 0.05em minus 0.05em 
    \textsf{\textcolor{black}{\small{\fontfamily{cmss}\selectfont ``#1''}}}%
    \endgroup
}
\newcommand{\camera}[1]{\textcolor{black}{#1}}

\newenvironment{widequote}
  {\begin{list}{}{%
    \setlength{\leftmargin}{0.8em}%
    \setlength{\rightmargin}{0.8em}%
  }\item\relax\small\itshape}
  {\end{list}}
  
\usepackage{xcolor}

\definecolor{objectTag}{RGB}{40,105,180}
\definecolor{actionTag}{RGB}{210,110,25}
\definecolor{closeTag}{RGB}{100,100,100}

\newcommand{\objecttag}[1]{%
  {\textcolor{objectTag}{%
    \footnotesize\ttfamily\strut\detokenize{#1}}}%
}

\newcommand{\actiontag}[1]{%
  {\textcolor{actionTag}{%
    \footnotesize\ttfamily\strut\detokenize{#1}}}%
}

\newcommand{\objecttext}[1]{%
  \begingroup
  \setlength{\fboxsep}{1.5pt}%
  \colorbox{objectTag!12}{\strut #1}%
  \endgroup
}

\newcommand{\actiontext}[1]{%
  \begingroup
  \setlength{\fboxsep}{1.5pt}%
  \colorbox{actionTag!15}{\strut #1}%
  \endgroup
}

\newcommand{\tagref}[1]{%
  \nobreak\textsuperscript{%
    \normalfont\scriptsize\textcolor{gray}{[#1]}%
  }%
}

\newcommand{\timenote}[3]{%
  \par\noindent
  \hangindent=2.2em
  \hangafter=1
  {\small\normalfont\color{gray}%
    \makebox[2.2em][l]{[#1]}%
    \texttt{#2}: #3%
  }%
}
\usepackage{microtype}
\newcommand{\tighttt}[1]{%
  {\small\texttt{#1}}%
}

\usepackage{comment}
\usepackage{ifthen}
\usepackage{xspace}
\usepackage{geometry}
\usepackage{graphicx}
\usepackage{array}
\usepackage{booktabs}
\usepackage{enumitem}

\usepackage{xcolor}

\AtBeginDocument{%
  }

\copyrightyear{2026}
\acmYear{2026}
\setcopyright{cc}
\setcctype{by}
\acmConference[UIST '26]{The 39th Annual ACM Symposium on User Interface Software and Technology}{November 02--05, 2026}{Detroit, MI, USA}
\acmBooktitle{The 39th Annual ACM Symposium on User Interface Software and Technology (UIST '26), November 02--05, 2026, Detroit, MI, USA}
\acmDOI{10.1145/3830398.3830476}
\acmISBN{979-8-4007-2856-3/2026/11}

\def\name{Sidekick}
\begin{document}

\title{{\name}: Designing Communication for Effective Multitasking with Computer Use Agents}

\author{Ruei-Che Chang}
\affiliation{
 \institution{UMich}
 \city{Ann Arbor, MI}
 \country{USA}
}
\email{rueiche@umich.edu}

\author{Wenqian Xu}
\affiliation{
 \institution{UMich}
 \city{Ann Arbor, MI}
 \country{USA}
}
\email{wxtu@umich.edu}

\author{Dingzeyu Li}
\affiliation{
 \institution{Adobe Research}
 \city{Seattle, WA}
 \country{USA}
}
\email{dinli@adobe.com}

\author{Bryan Wang}
\affiliation{
 \institution{Adobe Research}
 \city{Seattle, WA}
 \country{USA}
}
\email{bryanw@adobe.com}

\author{Anhong Guo}
\affiliation{
 \institution{UMich}
 \city{Ann Arbor, MI}
 \country{USA}
}
\email{anhong@umich.edu}

\begin{teaserfigure}
  \includegraphics[width=\textwidth, alt={}]{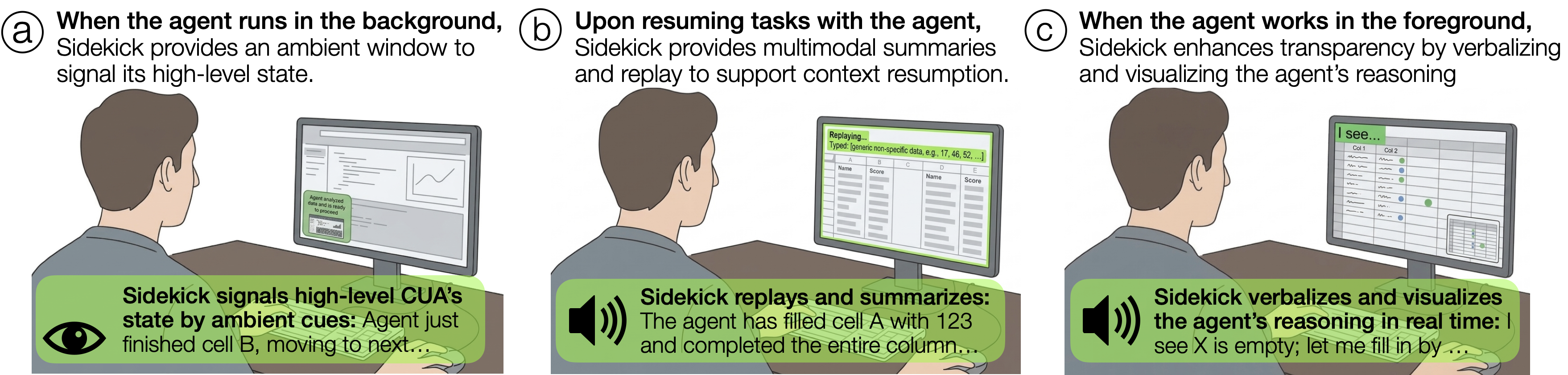}
  \vspace{-1.7pc}
  \caption{{\name} supports effective multitasking with computer-use agents (CUAs) by bridging communication gaps across different stages of interaction. In this example, CUAs assist with data annotation for a paper the user is writing.
  (a) When \emph{CUAs run in the background}, {\name} signals their execution states through ambient cues, mitigating disruption to the user’s primary task. 
(b) When \emph{the user resumes interaction}, {\name} presents multimodal summaries of completed actions, enabling rapid context resumption.
(c) When \emph{CUAs operate in the foreground}, {\name} improves their transparency by verbalizing and visualizing the agent’s reasoning process.
  }
  \Description{The figure consists of three horizontally arranged panels labeled (a), (b), and (c), each showing a person seated at a desk using a desktop computer, illustrating different interaction modes of an agent called ``Sidekick.''
In panel (a), titled When the agent runs in the background,'' the accompanying text reads: Sidekick provides an ambient interface to signal its state.'' The computer monitor displays a general interface with a small highlighted notification box. At the bottom, a green rectangular box with an eye icon contains the text: ``Sidekick signals high-level CUA's state by text: Agent just finished cell B, moving to next...'' This panel conveys that when the agent operates in the background, it provides passive, high-level textual updates about its progress.
In panel (b), titled Upon resuming tasks with the agent,'' the text reads: Sidekick provides multimodal summaries and replay to support context resumption.'' The monitor shows a spreadsheet-like interface with a green highlighted header area. At the bottom, a green rectangular box with a speaker icon contains the text: ``Sidekick replays and summarizes: I have filled column A by typing...'' This panel illustrates that when the user returns to a task, the system provides summaries and replays of prior actions to help restore context.
In panel (c), titled When the agent works in the foreground,'' the text reads: Sidekick enhances transparency by verbalizing and visualizing the agent's reasoning.'' The monitor displays a table with highlighted cells and a label at the top that reads I see...'' At the bottom, a green rectangular box with a speaker icon contains the text: Sidekick describes its reasoning in real time: I see X is empty, let me fill in by...'' This panel shows that when the agent is actively working in the foreground, it explains its reasoning step by step in real time.
In all three panels, the same user is depicted from behind, seated at a desk with a keyboard and mouse, facing a monitor. The use of green caption boxes and icons differentiates the communication modes: an eye icon for passive visual status signaling in panel (a), and speaker icons for spoken or narrated explanations in panels (b) and (c).
}
  \label{fig:teaser}
  \vspace{0pc}
\end{teaserfigure}

\begin{abstract}
Computer Use Agents (CUAs) can autonomously execute complex, multi-step tasks within GUIs, enhancing efficiency through parallel multitasking. However, our formative studies with CUA experts and GenAI users indicated that current feedback is primarily text-based, requiring sustained attention to monitor progress and offering limited visibility to trace past GUI interactions. Based on the findings, we developed a prototype system, Sidekick, for communicating CUAs’ status with multimodal feedback across different stages of interaction: 
\textit{(i)} When \textbf{CUAs run in the background}, Sidekick signals its execution state through ambient cues. 
\textit{(ii)} \textbf{Upon resuming interaction with CUAs}, Sidekick provides multimodal summaries of completed actions to support rapid context resumption. 
\textit{(iii)} When \textbf{CUAs operate in the foreground}, Sidekick enhances transparency by verbalizing and visualizing the agent’s reasoning. 
A study with 30 participants demonstrated that Sidekick significantly improved multitasking performance with CUAs compared to baseline systems that presented textual feedback either in a typical chat or in an ambient display. Sidekick supported progress awareness, and error and action traceability more effectively. 
Finally, we demonstrate the promise of Sidekick through several example applications, and discuss implications for long-horizon human-agent collaboration.
\end{abstract}

\ccsdesc[500]{Human-centered computing~Graphical user interfaces}
\ccsdesc[500]{Human-centered computing~User interface management systems}
\ccsdesc[500]{Human-centered computing~Auditory feedback}
\ccsdesc[500]{Human-centered computing~Information visualization}
\keywords{Computer use agent, multimodal feedback, ambient display, LLM, VLM, multitasking, human-agent collaboration}

\maketitle

\section{Introduction}
Computer Use Agents (CUAs) operate through sequences of actions within graphical user interfaces (GUIs), making intermediate decisions, interacting across applications, and executing tasks on users’ behalf.
These capabilities hold significant promise for autonomously managing complex tasks~\cite{yang2025ultracua, sager2026}, where users can delegate and coordinate tasks according to their preferences. 
However, CUAs may operate for extended periods, during which users often switch to other tasks. 
In certain situations, such as those involving privacy-sensitive or high-stake decisions~\cite{sager2026, Cheng_2026, Morae}, CUAs may still require timely user input. 
As a result, users must manage multiple tasks concurrently while maintaining effective communication and oversight of CUAs to ensure successful task execution.
This raises a key question: 
\emph{How should feedback be designed to support effective communication during multitasking with CUAs?}

To address this question, we conducted a formative study with three CUA experts and fifteen GenAI tool users (e.g., text, image, code generation). 
This allowed us to examine how issues observed in existing GenAI tools, which similarly involved AI reasoning, limited transparency, and extensive outputs, may transfer to CUAs, while grounding our analysis in CUA expert insights. 
We found that CUAs are often used in two interaction modes: \textbf{foreground}, where users actively monitor the agent’s progress, and \textbf{background}, where agents run autonomously while users attend to other tasks.
Across both modes, we identified several communication gaps.

When CUAs operate \textbf{in the foreground}, unlike other GenAI tools that display ongoing reasoning or real-time edits (e.g., vibe-coding tools), they provide limited transparency into their internal decision-making process and little GUI-visible feedback to support user engagement~\cite{Cheng_2026}.
When running \textbf{in the background}, CUAs also lacked clear signals of progress or completion. 
As a result, they often remain detached from users’ workflows and are checked asynchronously long after they are stuck or done.
\textbf{Upon resuming interaction with CUAs}, users encountered feedback that remains largely text-based, similar to other GenAI systems. 
This made it difficult to reconstruct CUAs' progress, intervene in time, or understand how GUI states have evolved, particularly when CUAs perform consequential actions and produce artifacts beyond text.

To bridge this communication gap in multitasking with CUAs, we introduce {\name}, a prototype system that supports monitoring and communicating CUA status across interaction stages: 
\textit{(i)} When CUAs operate \textbf{in the background}, {\name} conveys execution state with a high-level summary and visual thumbnail, along with ambient color cues in a peripheral display (e.g., green for normal progress, yellow for potential issues, red for being stuck).
\textit{(ii)} \textbf{Upon resuming interaction with CUAs}, {\name} provides multimodal summaries that combine speech and visual highlights of key actions and system changes to support rapid context recovery.
\textit{(iii)} When CUAs operate \textbf{in the foreground}, {\name} enhances transparency by verbalizing the agent’s reasoning in real time and visualizing its actions (e.g., annotated clicks, typing, and screenshots).

To evaluate {\name}, we conducted a mixed-methods study with 30 participants performing arithmetic questions as the primary task and spreadsheet filling as the secondary task. 
We found that {\name} significantly improved multitasking performance compared to users working alone, and two baseline systems that presented textual feedback either in a typical chat or on a peripheral display. 
This improvement was driven by better outcomes in the CUA-assisted secondary task, where {\name} enhanced users’ awareness of the CUA's states and enabled timely intervention when errors occurred without disrupting users’ primary tasks. 
Moreover, {\name} reduced errors more effectively than the baseline systems by providing multimodal summaries of CUAs' completed actions upon users' return, which were perceived as more helpful for understanding and multitasking with CUAs.
Finally, we demonstrated the promise of Sidekick through three application scenarios and discussed implications for long-horizon human-agent collaboration.

In summary, our work makes the following contributions:
\vspace{-1pc}
\begin{itemize}
\item[\textit{(i)}] A formative study with 15 GenAI users and 3 CUA experts that identifies key communication gaps between humans and CUAs during multitasking.

\item[\textit{(ii)}] {\name}, a prototype system that bridges these gaps by providing feedback to support transparency, progress awareness, and context resumption in human-CUA multitasking.

\item[\textit{(iii)}] A mixed-methods evaluation with 30 participants showing that {\name} significantly improved multitasking performance with CUAs compared to chat-based or peripheral-text feedback without disrupting the primary task.

\item[\textit{(iv)}] Design implications and three application scenarios to illustrate how {\name} may be extended to future long-horizon human-agent collaboration.
\end{itemize}
\section{Related Work}
Our work builds on prior research in 
\textit{(i)} computer-use and web agents that operate GUIs,
\textit{(ii)} transparency in human-agent collaboration, and
\textit{(iii)} ambient and peripheral displays that support multitasking, techniques for rapid context resumption.

\subsection{Computer-Use and Web Agents}
Recent advances in computer-use models have enabled agents that execute user goals by directly interacting with GUIs, bridging natural language intent and low-level actions (e.g., clicking, typing). Early work explored grounding models in web environments using screenshots, DOM structures, and action histories. For example, WebArena~\cite{webarena} and Mind2Web~\cite{deng2023mind2web} introduced realistic web environments and benchmarks that map high-level prompts to executable interactions in standardized settings. Building on these, WebVoyager~\cite{webvoyager} and SeeAct~\cite{seeact} leveraged large multimodal models (LMMs) to support more open-ended web tasks.

Beyond web agents, ``computer-use'' (OS-level) agents generalize these capabilities to desktop environments. Given a user goal, they iteratively infer executable actions from screenshots in a closed loop grounded in GUI states. 
Industry systems (e.g., OpenAI Operator~\cite{operator}, Anthropic Claude~\cite{claudeCUA}, Google Gemini~\cite{geminiCUA}) further demonstrate increasingly human-like interaction.
Correspondingly, evaluation has shifted from static, outcome-based metrics to \emph{human-centric}, trajectory-aware criteria that assess whether actions are visually grounded and contextually appropriate, which better captured real-world ambiguity and long-horizon tasks~\cite{gou2025mind2web, osworld, osworldhuman, windowsAgentArena}.

These advances suggest that CUAs are becoming increasingly autonomous, capable of handling complex tasks at scale~\cite{yang2025ultracua, sager2026}. However, increased autonomy also introduces risks, including errors and guardrails triggered by high-stakes actions~\cite{sager2026, Morae, openaiCUA}, which still require timely user intervention.
This raises a key question for the HCI community: \emph{How should feedback be designed to support effective communication during multitasking with CUAs?}
To address this, we conducted a formative study to derive design goals for {\name} to support human-CUA multitasking.

\subsection{Human-Agent Collaboration}
Principles for mixed-initiative systems emphasized coupling automation with direct manipulation, allowing users to invoke, override, or terminate automation~\cite{horvitz1999}. Human-AI guidelines similarly highlighted transparency, user correction, and error recovery~\cite{Amershi2019}.
Prior work shows transparency can improve performance and trust calibration, depending on task demands and interface integration~\cite{selkowitz2017using, van2024agent, vossing2022designing}. Accordingly, explanation techniques in different modalities, such as textual (what, why, consequences)~\cite{chen2018situation, EditScribe} and visual (e.g., feature highlights~\cite{lime}, interactive visualizations~\cite{whatif, ModelTracker, ExplainExplore, TimberTrek, VIME}), have been developed to support human-AI collaboration~\cite{Szymanski2021, TeleGam}.
However, these approaches are underexplored for CUAs, which operate over long horizons and may hallucinate, get stuck, or cause irreversible errors, which require timely communication for user intervention~\cite{Cheng_2026, Morae, mohanbabu2026a11y}. 
To address this gap, {\name} extends prior work by verbalizing and visualizing CUA reasoning and actions to enhance transparency in multitasking. Our study findings further highlight the need for context-aware, customizable communication to support effective human-CUA collaboration.

\subsection{Ambient Display, Multitasking, and Context Resumption}
HCI research on multitasking examined how users coordinated attention and managed interruptions. Prior work showed that interruption costs vary with timing~\cite{adamczyk_bailey2004} and task complexity~\cite{mcfarlane2002scope}. Several systems have been developed to mitigate disruption by inferring cognitive load and scheduling notifications opportunistically~\cite{Oasis, Iqbal2008, BusyBody, Lilsys, Chen2026Opportune, SoundShift}, or by supporting resumption through structured activity histories and video replay~\cite{TaskTracer, Chronicle}.
Another line of work explored awareness without undue distraction. 
Peripheral and ambient displays presented ``at-a-glance'' information in the periphery, enabling background monitoring while maintaining focus~\cite{ambient_taxonomy}. These systems provided interfaces spanning on-screen sidebars~\cite{side_bar_ambient}, multi-monitor setups~\cite{ambient_help, matejka2021meetingmate}, projections~\cite{kimura2001}, and physical artifacts with ambient cues~\cite{physical_ambient, Consolvo2005}. 
Prior studies also compared presentation strategies in ambient displays, including text vs.\ graphics~\cite{somervell2002evaluating}, motion vs.\ static~\cite{Maglio2000}, information density~\cite{dabbish_kraut2004}, and modality~\cite{Brewster1994, arroyo2002interruptions}.
These insights informed our design of ambient displays to unobtrusively signal CUAs’ states in {\name} and the baseline systems we compared in the evaluation (e.g., text in either chat or peripheral). 
{\name} builds on prior work in structured history and video replay to support resumption for CUAs that produce \emph{evolving GUI states}, rather than static or predetermined interfaces.

\section{Formative Study}\label{3_study}
In this study, we aimed to investigate how users interact with CUAs and GenAI tools across prompting, waiting, and result retrieval, and what feedback supports such collaboration.
\camera{We conducted a survey with 15 GenAI tool users (F1–F15; mean age = 26.5, SD = 3.3). The survey covered participants’ experiences with text, image, and code generation, with a focus on workflow integration, sustained engagement, and how users revisit and interpret generated outputs. We also conducted 30–60-minute contextual inquiries with three CUA experts (F16–F18; mean age = 29.3, SD = 3.2), during which they demonstrated their typical workflows and shared observations about CUA use. F16 used the CUA product developed by Vercept~\cite{vercept}, which was later acquired by Anthropic~\cite{anthropic}, for background tasks such as exploring and summarizing new papers and generating weekly stock charts. F17, a postdoctoral researcher, studied CUAs in assistive contexts, while F18, a PhD student, examined their use on complex cloud-based platforms. Collectively, their work spanned a range of CUA models and application contexts.}

This mixed-sample approach reflects the relatively early stage of CUA development and adoption. As of 2025, available CUAs remained limited~\cite{claudeCUA, openaiCUA, geminiCUA} and achieved approximately 60\% accuracy even on basic tasks~\cite{osworld, osworld-github}, constraining opportunities for broader real-world use. Our approach therefore allowed us to examine how challenges observed in established GenAI tools may extend to CUAs while grounding these insights in expert practice.

\begin{figure*}[t]
\begin{center}
\includegraphics[width=\linewidth]{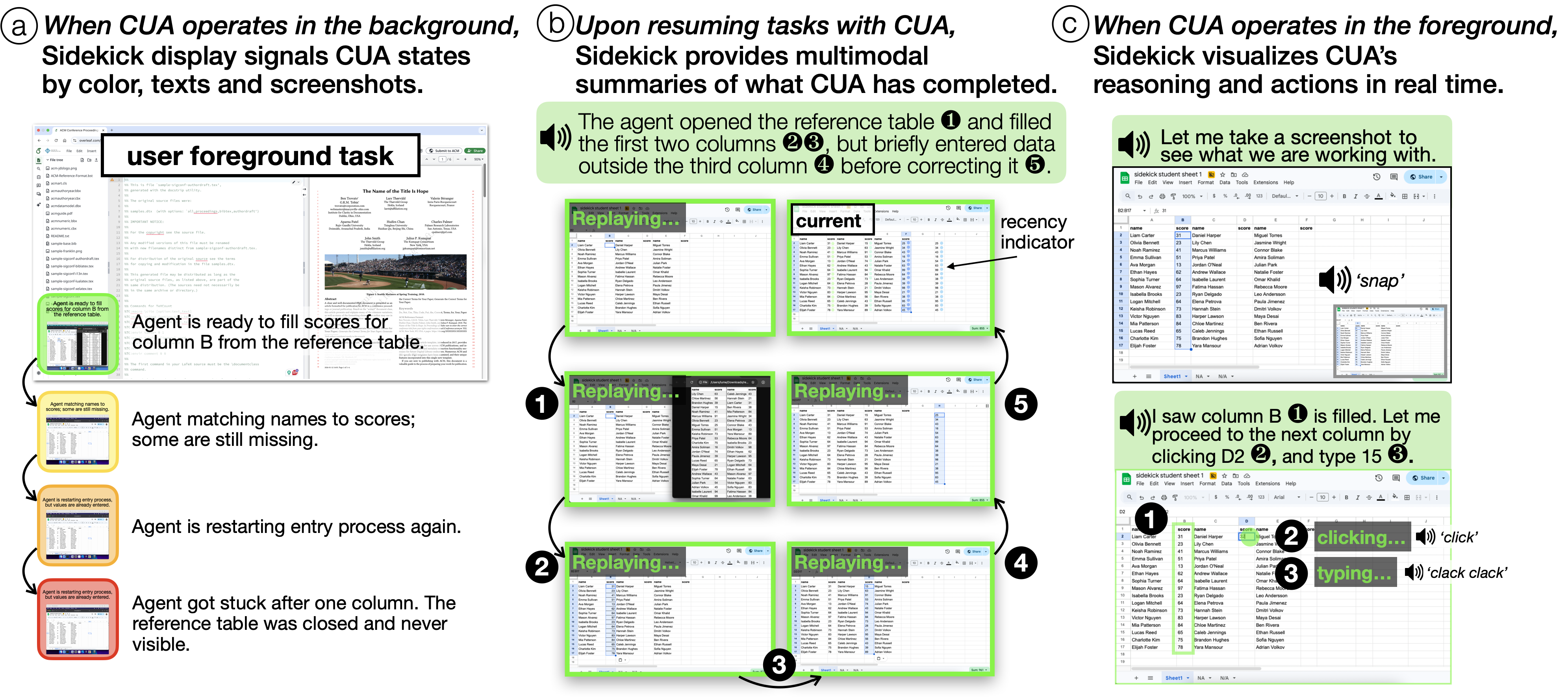}
\vspace{-1.8pc}
\caption{{\name}'s multimodal feedback across different stages of user interactions with CUAs.
(a) When the CUA runs in the background, {\name} provides peripheral awareness through ambient visual cues (colors, text, and thumbnails). Green indicates smooth progress, yellow signals minor errors, orange reflects accumulating issues, and red indicates the agent is stuck and has been paused for user intervention.
(b) Upon resuming, {\name} delivers multimodal summaries combining speech and visual replay, highlighting completed actions as well as encountered errors.
(c) When the CUA is in the foreground, {\name} visualizes the agent’s reasoning and actions in real time, augmented with synchronized speech and foley sound effects.
}
\label{fig:sidekick}
\Description{The figure is composed of three panels arranged horizontally and labeled (a), (b), and (c), each illustrating how a system called ``Sidekick'' supports a computer-using agent (CUA) in different modes of operation.
In panel (a), titled When CUA operates in the background,'' the text states: Sidekick displays signals CUA states by color, texts and screenshots.'' The main screen shows a document editing interface labeled user foreground task.'' To the left of the screen, there is a vertical sequence of four small thumbnail screenshots, each outlined with a different color (green, yellow, orange, and red), representing different agent states. Next to these thumbnails are textual descriptions: Agent is ready to fill scores for column B from the reference table,'' Agent matching names to scores; some are still missing,'' Agent is restarting entry process again,'' and ``Agent got stuck after one column. Something was closed and never visible.'' These elements indicate that Sidekick provides passive visual and textual updates about the agent’s background progress using color-coded signals and brief status messages.
In panel (b), titled Upon resuming tasks with CUA,'' the text reads: Sidekick provides multimodal summaries of what CUA has completed.'' At the top, a green speech box with a speaker icon contains the text: The agent opened the reference table (1) and filled the first two columns (2, 3), but briefly entered data outside the third column (4) before correcting it (5).'' Below this, a sequence of five screenshots is arranged in a flow, each labeled Replaying...'' except for one labeled current.'' The screenshots depict spreadsheet-like tables, and arrows connect them in order, with numbered markers (1 through 5) indicating the progression of actions. A label recency indicator'' points to the current state. This panel shows that Sidekick reconstructs and summarizes prior actions through both narration and visual replay to help the user regain context.
In panel (c), titled When CUA operates in the foreground,'' the text states: Sidekick visualizes CUA’s reasoning and actions in real time.'' At the top, a green speech box with a speaker icon reads: Let me take a screenshot to see what we are working with.'' Below it, a spreadsheet interface is shown, accompanied by the text snap,'' indicating a screenshot action. Further down, another green speech box reads: I saw column B (1) is filled. Let me proceed to the next column by clicking D2 (2), and type 15 (3).'' The spreadsheet display highlights specific cells with numbered markers (1, 2, and 3), and labels such as clicking...'' and typing...'' appear alongside sound annotations like click'' and ``clack clack.'' This panel demonstrates that when the agent is actively working, Sidekick provides step-by-step explanations of its reasoning and actions, synchronized with visual highlights and interaction cues.
}
\vspace{-1pc}
\end{center}
\end{figure*}

\subsection{Results}\label{3_results}
\subsubsection{Some users treated GenAI tools as peripheral, background components in multitasking workflows.}\label{3_genai_pattern}
Users integrated GenAI tools into broader workflows alongside productivity tools (e.g., image generation for slides in PowerPoint or Keynote, and text generation for writing in Overleaf). 
They often used multiple GenAI tools concurrently to support multitasking.
As a result, GenAI tools were typically treated as peripheral or background components, running in browser tabs, split-screen layouts, or small floating windows rather than occupying primary focus. In contrast, code generation tools were more tightly embedded within development environments (e.g., Cursor~\cite{cursor}, Claude Code~\cite{claude}).
Similarly, CUAs were frequently executed in the background or in the cloud for scheduled or long-running tasks to avoid interfering with users’ local computer and primary work (e.g., F16 used Vercept~\cite{vercept} to generate weekly stock market plots in Google Sheets).

\subsubsection{Some users shifted attention to manually monitor progress}\label{3_attention_switch}
When tasks were complex or time-consuming, some users treated GenAI tools as background processes and shifted to other tasks (N=12), such as during \user{refactoring or making big changes} (F3). Because intermediate progress signals were limited, participants periodically checked progress by using Alt+Tab (F11, F12), switching back \user{from time to time} (F2), or monitoring screens and terminal messages (F17). Although disruptive, these checks helped users maintain context and avoid the cost of resuming interrupted work, as F5 explained: \user{I would constantly check the progress because I don't want to lose my flow.}
These findings highlight the need for clearer, real-time indications of GenAI progress.

\subsubsection{Some users sustained attention to GenAI tools}\label{3_sustained_attention}
Some users remained attentive to GenAI tools when reviewing outputs or handling high-stakes tasks. F8 stayed on ChatGPT \user{for important tasks} to pause or revise prompts. Others monitored only the initial output before returning to their main work; for example, F13 verified Claude’s to-do list before switching away.
Participants also used secondary monitors (F10) or split-screen layouts (F13) to maintain peripheral awareness. Fast inline feedback in GenAI tools further reduced attention shifts, as F4 noted: \user{The inline suggestions appear quite quickly, so that I can use them immediately without shifting.} However, feedback in current CUA tools was often overwhelming, whether presented as dense terminal text (F17, F18) or as a combination of screenshots, actions, and reasoning (F16). These findings highlight the need for timely, unobtrusive feedback that supports lightweight monitoring.

\subsubsection{Current interface features to support task switching are insufficient}\label{3_current_genai_features}
To support task switching, particularly when returning to GenAI tools, participants relied on notifications to signal completion. However, prolonged waiting made re-engagement difficult, often requiring users to scroll through logs or chat histories to reconstruct context. As F10 noted, lengthy responses made it easy to \user{lose track of where the conversation is going.}
Interface structuring partially mitigated this issue. Elements such as \emph{code blocks}, \emph{bold headings}, and clear response semantics helped users reorient; for example, F3 described scanning high-level explanations before examining detailed edits.
However, current CUA chat interfaces offer limited structured or layered feedback, making users rely primarily on final screen states or text-based reasoning.
This highlights the need for better support for task resumption, especially as CUAs produce consequential actions beyond text.

\section{{\name}}
{\name} is a prototype system designed to bridge the communication gap across different interaction stages with CUAs. 
Based on the formative study findings, we derived three design goals: \begin{itemize}[leftmargin=1.8em]
    \item[\textbf{D1}] \textit{When CUAs operate in the background}: Enable peripheral awareness of CUAs' intermediate progress to reduce manual checking (Sections~\ref{3_genai_pattern} \& \ref{3_attention_switch}).
    \item[\textbf{D2}] \textit{Upon resuming tasks with CUAs}: Offer structured, high-level summaries of completed actions, system changes to support rapid context resumption (Sections~\ref{3_attention_switch} \& \ref{3_current_genai_features}).
    \item[\textbf{D3}] \textit{When CUAs operate in the foreground}: Provide real-time feedback on CUAs' actions and reasoning to support transparency (Section~\ref{3_sustained_attention}).
\end{itemize}
Specifically, it provides peripheral signals to convey task progress when running in the background (Section~\ref{4_system_background}), structured multimodal summaries to communicate completed actions and changes (Section~\ref{4_system_transition}), and real-time, synchronized text and visual feedback when CUAs operate in the foreground (Section~\ref{4_system_foreground}).
Below, we illustrate these {\name} interactions across stages using an example in which a user writes a paper while a CUA fills a spreadsheet with relevant data (Figure~\ref{fig:teaser}).

\subsection{CUAs operating in the background}\label{4_system_background}
To fulfill \textbf{D1} and support peripheral awareness of CUA progress, {\name} introduces an adjustable, always-on-top peripheral window, referred to as the \textit{{\name} display}. The display communicates high-level CUA activity through concise text summaries, screenshots, and color-coded status signals (Figure~\ref{fig:sidekick}a).
Within this window, {\name} condenses raw CUA messages into brief, one-sentence summaries (e.g., \agent{Agent found data references, is analyzing the spreadsheet, and will fill Column B starting at B2}). These summaries allow users to monitor progress at a glance while remaining focused on their primary task. For later reference, the same message also appears in the chat interface (Figure~\ref{fig:setup}d, shown in black).
\camera{{\name} evaluates each generated action by comparing screenshots taken before and after execution. A VLM (\tighttt{gemini-2.5-flash}) assesses whether the resulting state matches the user’s query or the CUA’s current subgoal. An action is marked erroneous when the post-execution state fails to reflect the intended outcome. Most errors involved either \textit{no progress}, where the interface remained unchanged, or an \textit{incorrect action}, where the resulting state diverged from the goal.}
Based on this definition, {\name} tracks the number of consecutive errors, denoted by $e$, and adjusts the color of the \textit{{\name} display} to indicate system status: yellow when $e > x$, orange when $e > y$, and red when $e > z$. Upon reaching the red state, {\name} automatically pauses the CUA as a safeguard. This design aims to draw the user’s attention when intervention may be necessary while remaining minimally intrusive during normal operation. In our study, we set $x=3$, $y=6$, and $z=8$ based on the error patterns observed during prototype development.

\subsection{Resuming tasks with CUAs}\label{4_system_transition}
To address \textbf{D2} and support context resumption, {\name} generates a multimodal summary when users return to CUAs (e.g., focusing on the CUA's tasks), describing actions completed during background execution (Figure~\ref{fig:sidekick}b).
To achieve this, {\name} operates continuously in the background and produces summaries at each completed step by prompting a VLM with prior actions, step indices, screenshots, and agent messages.
The summaries are converted into the Speech Synthesis Markup Language (SSML)~\cite{ssml} with embedded step IDs, allowing text-to-speech (TTS) services to generate timestamps for synchronized playback. 
Using these, {\name} aligns narration with corresponding screenshots and action visualizations (Detailed in Section~\ref{4_system_implementation}).
Depending on user preference, summaries can be presented as either audio or text. \camera{The CUA remains paused during summary playback, allowing users to review the completed actions before execution resumes.}
After playback, {\name} visualizes action recency using a color gradient (e.g., green for earlier actions, blue for recent ones), helping users understand how prior steps led to the current system state.

\subsection{CUAs operating in the foreground}\label{4_system_foreground}
To fulfill \textbf{D3} and provide transparent, real-time feedback, {\name} externalizes the CUA’s reasoning, plans, and actions through multimodal cues when users actively engage with the agent (Figure~\ref{fig:sidekick}c). \camera{This includes verbal and visual feedback to accommodate different attention preferences and interaction contexts.}
Specifically, {\name} verbalizes the CUA’s messages, including internal planning (e.g., \agent{I can see column B is empty and needs to be filled. Now I need to reference the Preview app to get the correct scores.}) and action-completion updates (e.g., \agent{I’ve typed 65 in cell D7. Now I’ll press Enter to confirm the entry.}). 
It also generates Foley sounds~\cite{cheng2026auditorily, foley} to accompany computer interactions such as clicking, typing, and taking screenshots.
In parallel, {\name} visualizes actions on the screen, in sync with the speech and foley sounds. 
For example, it highlights spoken content with cropped bounding boxes (Figure~\ref{fig:sidekick}.1), annotates clicked locations with visual markers (Figure~\ref{fig:sidekick}.2), and displays the issued keyboard commands in texts (Figure~\ref{fig:sidekick}.3).
Similar to multimodal summaries, this synchronization is achieved by converting messages into SSML~\cite{ssml} with embedded action labels, enabling TTS to generate timestamps that align speech with visualizations. We detail the implementation in the next section.

\subsection{Implementation details}\label{4_system_implementation}
We use the existing CUA framework~\cite{cua} and the model \tighttt{anthropic/\allowbreak-claude\allowbreak-sonnet\allowbreak-4-5\allowbreak-20250929}. 
The CUA runs inside a macOS virtual machine (Figure~\ref{fig:setup}d), Lume~\cite{lume}, hosted on a local MacBook Air (Figure~\ref{fig:setup}a).
All click-through on-screen visualizations, including peripheral windows and annotations, are developed in Swift and deployed on both the macOS virtual machine (VM) and the MacBook host.
\camera{For the real-time multimodal feedback described in Sections~\ref{4_system_transition} and~\ref{4_system_foreground}, including synchronized verbalizations, visualizations, and foley sounds, we use \texttt{\small gemini-2.5-flash} as a VLM-based converter that transforms plain-text CUA messages into SSML-like annotations~\cite{ssml}. The VLM is prompted to identify references to GUI objects and actions in {\name}'s messages and ground them in the current screenshot. For example, the CUA message in one of the proposed application scenarios (Figure~\ref{fig:learnapp}):}
\begin{widequote}
\textit{``With the cat now selected with marching ants, the agent will click the generative fill button.''}
\end{widequote}
can be converted into the following SSML-like annotation:
\begin{widequote}
\textit{%
``With the
\objecttag{<mark name="object_1"/>}\tagref{1}\hspace{0.15em}%
\objecttext{cat}\hspace{0.15em}%
\objecttag{</>}\tagref{2}
now selected with marching ants, the agent will
\actiontag{<mark name="action_1"/>}\tagref{3}\hspace{0.15em}%
\actiontext{click}\hspace{0.15em}%
\actiontag{</>}\tagref{4}
\objecttag{<mark name="object_2"/>}\tagref{5}\hspace{0.15em}%
\objecttext{the generative fill button}\hspace{0.15em}%
\objecttag{</>}\tagref{6}.''
}

\medskip
\timenote{1}{0.3\,s}{Highlight the cat with a bounding box.}
\timenote{2}{1.1\,s}{Remove the cat bounding box.}
\timenote{3}{3.9\,s}{Display ``click'' and mark the click location.}
\timenote{4}{4.2\,s}{Remove the action cue.}
\timenote{5}{4.2\,s}{Highlight the button with a bounding box.}
\timenote{6}{5.6\,s}{Remove the button bounding box.}
\end{widequote}
\camera{
Object labels are sent to a VLM (\texttt{\small gemini-3.5-flash}) for open-vocabulary object detection, which returns bounding boxes for visualization, while action labels specify when to render on-screen action cues and Foley sounds. 
The SSML-like annotation is then passed to Google TTS~\cite{googletts}, which synthesizes speech and produces timestamps that {\name} uses to synchronize verbal, visual, and auditory feedback.
}
\section{Experiment}
To understand the effectiveness of {\name} in supporting multitasking with CUAs, we conducted a mixed-methods study with 30 participants, recruited via public posts at our institution (Mean age = 22.6, 18 female and 12 male), to quantitatively and qualitatively examine how {\name} supports multitasking with CUAs (Table~\ref{tab:demographic}).

\subsection{Tasks}
\begin{figure}[h]
\centering
\includegraphics[width=\linewidth]{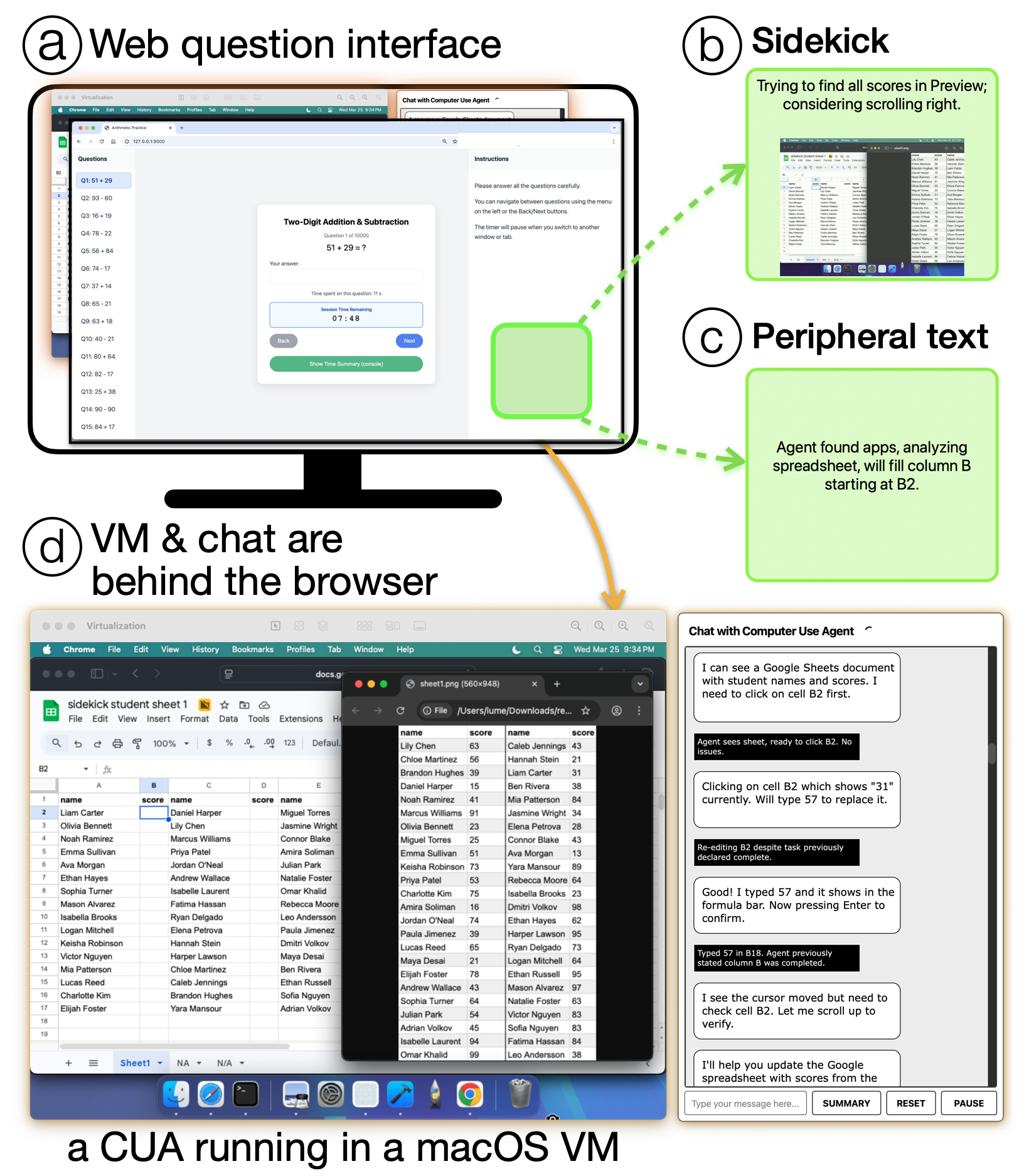}
\vspace{-1.8pc}
\caption{Study setup in a monitor.
(a) Participants interacted with a browser for arithmetic questions, accompanied by a peripheral display for (b) \textit{Sidekick} and (c) \textit{Peripheral Text} conditions.
(d) The CUA operated on a macOS virtual machine with its chat interface positioned behind the browser. 
}
\label{fig:setup}
\vspace{-1pc}
\Description{The figure is divided into four labeled components (a), (b), (c), and (d), illustrating a system setup involving a web interface, a Sidekick assistant, peripheral text, and a virtual machine (VM) environment.
In panel (a), labeled Web question interface,'' a desktop monitor displays a browser-based interface. On the left side of the interface is a vertical list of timestamps or entries (e.g., 02:15:30,'' 02:15:31,'' and so on). In the center, a modal dialog box titled Two-Clip Addition & Subtraction'' is visible, containing labeled input fields such as 1st number'' and 2nd number,'' with example values shown as 42 + 29.'' Below the inputs is a button labeled Solve this Sample Question.'' The right side of the interface contains instructional text beginning with ``Please answer all the questions carefully...'' A green dashed arrow extends from the monitor toward panel (b), indicating communication with the Sidekick system.
In panel (b), labeled Sidekick,'' a green box contains a small screenshot of a spreadsheet interface along with a textual message at the top that reads: Trying to find all scores in Preview; considering scrolling right.'' This represents the Sidekick providing contextual reasoning or status updates based on the agent’s actions.
In panel (c), labeled Peripheral text,'' another green box contains a textual summary: Agent found apps, analyzing spreadsheet, will fill column B starting at B2.'' A dashed green arrow connects this box to the main interface in panel (a), indicating that this peripheral text serves as an auxiliary explanation or status display.
In panel (d), labeled VM & chat are behind the browser,'' the lower portion of the figure shows a macOS desktop environment. On the left, a spreadsheet application is open, displaying columns such as name'' and score,'' with rows of names and numerical values. In the center, a smaller window shows another spreadsheet or reference table. On the right, a chat window titled Chat with Computer Use Agent'' contains a sequence of messages describing the agent’s actions, including: I can see a Google Sheets document with student names and scores. I need to click on cell B2 first,'' Clicking on cell B2 which shows '31' currently. Will type 57 to replace it,'' Good! I typed 57 and it shows in the formula bar. Now pressing enter to confirm,'' I see the cursor moved but need to check cell B2. Let me scroll up to verify,'' and I’ll help you update the Google spreadsheet with scores from the reference table.'' At the bottom of the chat window are buttons labeled SUMMARY,'' RESET,'' and PAUSE.'' A caption below the entire panel reads: ``a CUA running in a macOS VM.''
}
\end{figure}

To simulate multitasking, we carefully designed a primary and a secondary task based on prior work, pilot studies, and current CUA capabilities.
Participants solved arithmetic problems as the \textbf{primary task} (e.g., two-digit addition or subtraction) while delegating a spreadsheet-filling task to the CUA as the \textbf{secondary task} (e.g., entering scores from a reference table; Figure~\ref{fig:setup}d).
We chose two-digit arithmetic tasks as they were commonly cited as cognitively demanding tasks in multitasking studies~\cite{briem1995behavioural, czerwinski2000instant, rubinstein2001executive, Cheng2025agent}.
Participants were placed in a scenario in which they served as a teaching assistant entering scores into an outdated grading system while managing an arithmetic assignment deadline, motivating them to seek help from CUAs. 
We selected spreadsheet-filling tasks due to the limited capabilities of current CUAs (mentioned in Section~\ref{3_study}). Based on our empirical observations, CUAs were more controllable on these tasks and achieved higher success rates than on complex ones (e.g., slide or image editing).

For the secondary spreadsheet-filling task, the CUA was prompted to fill an entire column and then intentionally introduce two random errors to simulate unexpected behavior that could lead to serious mistakes (e.g., incorrectly failing students), requiring user attention.
This was repeated across three columns (16 entries each). Each correct entry earned 3 points, incorrect entries incurred a 12-point penalty, and blanks received 0.
Arithmetic answers were worth 1 point, with a 1-point penalty for errors and 0 for blanks.
The spreadsheet task was weighted three times higher because it took human users about three times longer to complete.
The 12-point penalty was carefully calibrated to encourage participants to monitor potential CUA errors: pilot studies showed that smaller penalties led participants to overlook errors, while larger penalties discouraged users from using CUA.

Note that {\name} is not an agent or CUA, but a prototype system that serves as a communication layer for users to multitask with CUAs. It does not perform actions on the computer, but only provides feedback to users. Future work may explore integrating error remediation agents alongside CUAs to address errors automatically, which is beyond the scope of this paper.

\subsection{Conditions and Research Questions}
We included four multitasking conditions in this study:
\begin{itemize}
\item[\textbf{(MN)}] \emph{Manual}: the user completes both tasks on their own.
\item[\textbf{(BL)}] \emph{Baseline}: a standard chat interface displaying only the CUA’s textual messages.
\item[\textbf{(PT)}] \emph{Peripheral Text}: a standard chat interface, and a peripheral display presenting high-level summaries of the CUA’s current states.
\item[\textbf{(SK)}] \emph{Sidekick}: a standard chat interface and a \textit{{\name} display}, as well as other multimodal feedback and summary.
\end{itemize}
We included \textbf{MN} to establish participants’ baseline multitasking performance and to examine whether CUAs could improve performance compared to humans working alone. 
We included \textbf{BL} because chat-based interfaces currently represent the most common medium of communication with CUAs in existing products (e.g., Gemini~\cite{geminiCUA}, Operator~\cite{operator}).
\textbf{PT} incorporated textual feedback presented in a peripheral display, reflecting the design of prior ambient systems.
Finally, \textbf{SK} included all {\name} features described in the previous section.
Across these four conditions, we investigated the following research questions:
\begin{itemize}

\item[\textbf{RQ1}] How do different feedback conditions affect users’ multitasking performance when collaborating with a CUA?

\item[\textbf{RQ2}] How do users allocate attention and interact with the CUA under different feedback conditions during multitasking?

\item[\textbf{RQ3}] How do different feedback conditions influence users’ perceived cognitive load, trust, and overall experience of multitasking with CUAs?

\end{itemize}

\begin{figure}[b]
\vspace{-1pc}
\centering
\includegraphics[width=\linewidth]{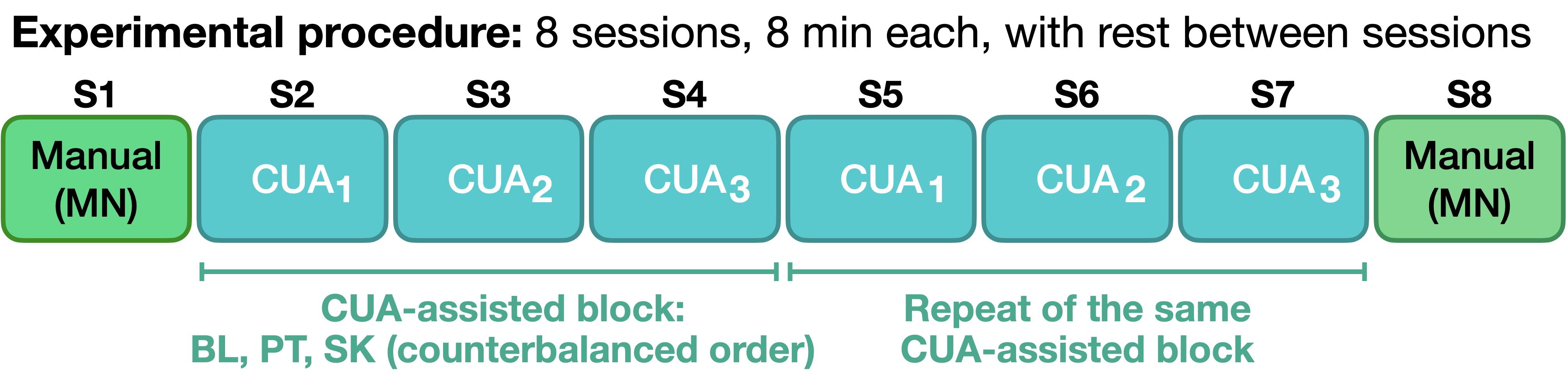}
\vspace{-1.8pc}
\caption{Experimental procedure. Participants completed eight 8-minute sessions: manual, two counterbalanced CUA-assisted blocks, and a final manual session. 
}
\label{fig:procedure}
\Description{The image presents a horizontal study procedure on a black background. It begins with a green rounded box labeled “Manual (MN),” followed by three turquoise boxes labeled “CUA1,” “CUA2,” and “CUA3.” These three boxes form a CUA-assisted block consisting of BL, PT, and SK in a counterbalanced order. The same sequence of CUA1, CUA2, and CUA3 is then repeated as a second CUA-assisted block, and the procedure ends with another green rounded box labeled “Manual (MN).” Overall, the sequence is Manual, CUA1, CUA2, CUA3, CUA1, CUA2, CUA3, and Manual.}
\end{figure}

\begin{figure*}[t]
\begin{center}
\includegraphics[width=\linewidth]{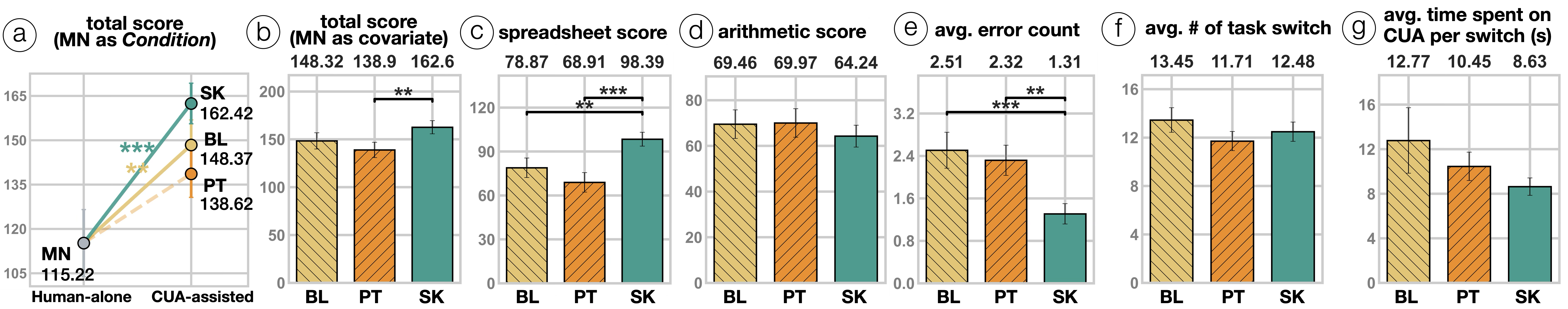}
\vspace{-1.8pc}
\caption{
\camera{
Results across conditions (\textbf{MN}, \textbf{BL}, \textbf{PT}, and \textbf{SK}) for task performance and interaction metrics. Panel (a) presents results from analyses treating \textbf{MN} as a fixed-effect condition alongside \textbf{BL}, \textbf{PT}, and \textbf{SK}. The remaining panels present results from analyses treating \textbf{MN} as a covariate: (b) \textit{total score}, (c) \textit{spreadsheet score}, (d) \textit{arithmetic score}, (e) \textit{average error count}, (f) \textit{average number of task switches}, and (g) \textit{average monitoring time per switch} for the CUA, in seconds. The mean scores in panels (a) and (b) slightly differ because the analyses modeled \textbf{MN} as a condition in panel (a) and as a covariate in panel (b).
Bars represent means, and error bars indicate $\pm 1$ standard error.
Statistical significance is denoted by *$p<.05$, **$p<.01$, and ***$p<.001$.}
}
\label{fig:results}
\vspace{-1pc}
\Description{The figure contains seven panels labeled a through g, comparing the BL, PT, and SK conditions. Panel a is a line plot of total score with Manual, or MN, treated as a condition: all three colored lines begin at the MN score of 115.22 and rise to PT at 138.62, BL at 148.37, and SK at 162.42; significance markers show two asterisks for BL and three asterisks for SK. Panel b shows total score with MN treated as a covariate: BL is 148.32, PT is 138.9, and SK is 162.6, with a significant difference between PT and SK marked by two asterisks. Panel c shows spreadsheet scores of 78.87 for BL, 68.91 for PT, and 98.39 for SK; SK differs significantly from BL with two asterisks and from PT with three asterisks. Panel d shows arithmetic scores of 69.46 for BL, 69.97 for PT, and 64.24 for SK, with no significance markers. Panel e shows average error counts of 2.51 for BL, 2.32 for PT, and 1.31 for SK; SK has significantly fewer errors than BL, marked by three asterisks, and PT, marked by two asterisks. Panels f and g have no visible metric titles in the image: panel f shows values of 13.45 for BL, 11.71 for PT, and 12.48 for SK, while panel g shows 12.77 for BL, 10.45 for PT, and 8.63 for SK. All bar charts include error bars; BL is shown in pale yellow with diagonal hatching, PT in orange with diagonal hatching, and SK in teal.}
\end{center}
\end{figure*}

\subsection{Interface Setup and Requirements}
We provided participants with a MacBook Air running both {\name} and a CUA in real time. 
Arithmetic tasks were presented in a Google Chrome interface, where users navigated via a sidebar or pressed \textit{Enter} to submit answers and proceed (Figure~\ref{fig:setup}a), with a visible session timer.
A separate VM window displayed the CUA and its chat interface, which included three controls: “RESET” (restart without context), “PAUSE” (temporarily halt while preserving context), and “SUMMARY” (exclusive to \textbf{SK}, providing on-demand summaries of the CUA's previous actions).
The Chrome, VM and chat windows had fixed sizes, with the VM and chat positioned behind Chrome, and participants could not modify the layout. 
A movable peripheral display was available in \textbf{PT} and \textbf{SK} (Figure~\ref{fig:setup}b,c).
The VM window, peripheral display, and chat interface were linked: focusing on one brought the others to the foreground.
\camera{An OS-level script ran in the background to record how long each window remained in focus for further analysis (Figure~\ref{fig:results}f).}

\subsection{Procedure}
\camera{The study consisted of eight 8-minute sessions with sufficient rest provided between sessions (Figure~\ref{fig:procedure}).
Participants were first introduced to all four multitasking conditions, including \textbf{MN}, \textbf{BL}, \textbf{PT}, and \textbf{SK}, and practice their delegated tasks.}
Then, participants began with the \textbf{MN} condition, followed by the other six sessions with three CUA-assisted conditions, including \textbf{BL}, \textbf{PT}, and \textbf{SK}, repeating twice and ended with another \textbf{MN} session. 
The order of three CUA-assisted conditions was counterbalanced across participants. The first and last \textbf{MN} sessions served to measure baseline performance and potential fatigue or learning effects. 
\camera{Participants were encouraged to maximize their scores, although their compensation was independent of task performance.} After each session, participants completed Likert-scale questionnaires and the NASA-TLX~\cite{nasatlx} to assess cognitive load, after which they received feedback on their session scores. The study concluded with a post-study interview.

\subsection{Dependent Measures and Data Analysis}
For each session, we recorded task performance measures, including scores, error counts, task-switching behavior, and time spent across windows.
We fitted two linear mixed-effects models. First, to compare CUA-assisted conditions (\textbf{BL}, \textbf{PT}, \textbf{SK}) with manual performance (\textbf{MN}), we included \textit{Condition} as a fixed effect, \textit{trial order} as a covariate, and participant ID as a random intercept. 
Second, to compare CUA-assisted conditions only, we fitted a model with \textit{Condition} (\textbf{BL}, \textbf{PT}, \textbf{SK}) as a fixed effect and participant ID as a random intercept~\cite{lindstrom1988newton}. 
We also included \textit{trial order} and \textit{manual performance} (mean performance across the two \textbf{MN} sessions) as covariates to control for order effects and individual differences. 
\camera{Including \textit{manual performance} as a covariate helped account for baseline ability, reducing the risk that observed differences were driven by individual skill rather than system effects and better isolating system-level gains.}
We did not observe fatigue or learning effects between the first and second \textbf{MN} conditions using paired t-tests and Wilcoxon signed-rank tests ($p > .16$).
We did not include the first or second parts of the study as a covariate, as no significant effect was shown on performance ($\chi^2(1) = 1.15$, $p = .28$). 
Pairwise contrasts between conditions were tested using Wald tests with Holm correction for multiple comparisons.

\begin{figure*}[h]
\begin{center}
\includegraphics[width=\linewidth]{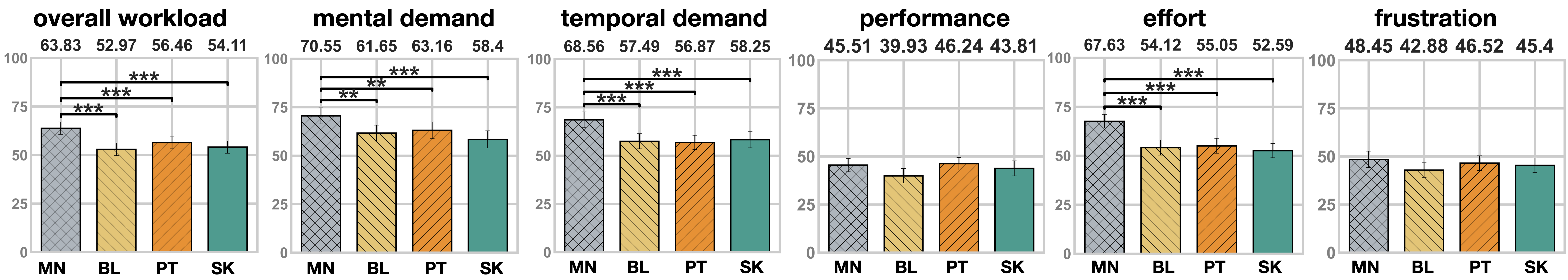}
\vspace{-1.8pc}
\caption{
NASA-TLX responses across conditions (\textbf{MN}, \textbf{BL}, \textbf{PT}, and \textbf{SK}) for \textit{overall workload}, \textit{mental demand}, \textit{temporal demand}, \textit{performance} (the higher, the better), \textit{effort}, and \textit{frustration}. We excluded \textit{physical demand} because it was not relevant to the study context. Bars show mean values with error bars indicating $\pm$1 standard error. Statistical significance is indicated as $p<.05$ (*), $p<.01$ (**), and $p<.001$ (***).
}
\label{fig:nasa}
\Description{The figure consists of six horizontally arranged bar charts titled overall workload,'' mental demand,'' temporal demand,'' performance,'' effort,'' and frustration.'' Each chart compares four conditions labeled MN, BL, PT, and SK. Numerical values are displayed above each bar, and statistical significance is indicated using asterisks (*, **, ***).
In the overall workload'' chart, the values are MN (63.83), BL (52.97), PT (56.46), and SK (54.11). MN has the highest workload, while BL has the lowest. Multiple significance markers are shown, including triple asterisks (***''), indicating significant differences between MN and the other conditions.
In the mental demand'' chart, the values are MN (70.55), BL (61.65), PT (63.16), and SK (58.4). MN again has the highest value, and SK the lowest. Both double ('') and triple (``*'') asterisks indicate statistically significant reductions in mental demand compared to MN.
In the temporal demand'' chart, the values are MN (68.56), BL (57.49), PT (56.87), and SK (58.25). MN is the highest, while PT is the lowest. Triple asterisks (***'') indicate significant differences between MN and the other conditions.
In the ``performance'' chart, the values are MN (45.51), BL (39.93), PT (46.24), and SK (43.81). PT has the highest value, followed by MN and SK, with BL the lowest. No statistical significance markers are shown in this panel.
In the effort'' chart, the values are MN (67.63), BL (54.12), PT (55.05), and SK (52.59). MN has the highest effort, and SK the lowest. Triple asterisks (***'') indicate significant reductions in effort for the other conditions compared to MN.
In the ``frustration'' chart, the values are MN (48.45), BL (42.88), PT (46.52), and SK (45.4). MN has the highest frustration, while BL has the lowest, with PT and SK in between. No significance markers are shown in this panel.
}
\end{center}
\end{figure*}

\subsection{Results}

\subsubsection{RQ1: How do different feedback conditions affect users’ multitasking performance when collaborating with a CUA?}\label{5_RQ1_results}\textbf{Human-CUA collaboration improves multitasking performance over human alone, and {\name}'s multimodal feedback further enhances this collaboration without disrupting users’ primary task.}
We evaluated multitasking performance using three metrics: \emph{total score}, \emph{arithmetic score}, and \emph{spreadsheet score}, where \emph{total score} is the sum of the other two (Figure~\ref{fig:results}). 
\camera{We first compared \textbf{MN} with CUA-assisted conditions (including BL, PT, and SK) to assess whether human-CUA collaboration improves \emph{total score} over a human alone (Figure~\ref{fig:results}a).
A linear mixed-effects model revealed a significant main effect of \emph{Condition} on \emph{total score} (Wald $\chi^2(3)=23.03$, $p<.001$). The estimated marginal means were 115.22 for \textbf{MN}, 148.37 for \textbf{BL}, 138.62 for \textbf{PT}, and 162.42 for \textbf{SK}. Holm-corrected pairwise comparisons showed that both \textbf{BL} ($p=.007$) and \textbf{SK} ($p<.001$) significantly outperformed \textbf{MN}. Although \textbf{PT} also achieved a higher score than \textbf{MN}, the difference was not significant after correction ($p=.075$), suggesting that CUAs are not necessarily beneficial when their feedback is not well designed.
}

\camera{We next compared \emph{total score} across the CUA-assisted conditions (\textbf{BL}, \textbf{PT}, and \textbf{SK}; Figure~\ref{fig:results}b). To account for individual differences in general task ability, we included each participant’s performance in \textbf{MN} as a covariate in the linear mixed-effects model.} A significant main effect of \emph{Condition} was observed (Wald $\chi^2(2)=12.31$, $p=.002$), with \textbf{SK} (M=162.6) outperforming \textbf{PT} (M=138.9, $p=.001$) and showing a marginal trend over \textbf{BL} (M=148.32, $p=.079$).

To understand the source of this improvement, we examined the two task components separately (Figure~\ref{fig:results}c, d).
We found that the effect of \emph{Condition} on \emph{total score} was driven by \emph{spreadsheet score} (Wald $\chi^2(2)=21.52$, $p<.001$), where \textbf{SK} (M=98.39) significantly outperformed both \textbf{PT} (M=68.90, $p<.001$) and \textbf{BL} (M=78.87, $p=.006$). In contrast, \emph{arithmetic score} showed no significant differences (Wald $\chi^2(2)=3.01$, $p=.222$). This indicated that the different feedback mechanisms used to communicate with the CUA did not interfere with users’ primary arithmetic task performance.

These results suggest that \emph{human-CUA collaboration enables users to achieve higher performance than human alone, but the communication mechanisms must be carefully designed.} 
For instance, \textbf{PT} did not outperform \textbf{BL}, as peripheral text and static color provided weak signals of the CUA’s state. 
Furthermore, the unchanging color might create a false sense of security, leading users to disengage and feel it \user{made it seem everything was fine} (P9). 
Also, text feedback required additional effort to interpret, \user{without color changes, it takes extra time to read the text and figure out if something is wrong} (P15), and was sometimes overlooked as \user{it was impossible to do the calculation and see the text at the same time} (P17).

In contrast, \emph{{\name}’s multimodal feedback improved collaboration without disrupting the primary task.}
However, preferences on multimodal feedback varied: some participants favored a simpler design (\textbf{BL}), noting that the rich information in \textbf{SK} could be distracting. 
As P10 noted, \user{Too many things going on split my attention and threw me off. I preferred the baseline for simplicity. I’d rather have a simple, on-demand color cue.} 
This attentional fragmentation may explain why \textbf{SK} showed only marginal improvement over \textbf{BL}.
We next analyze user behaviors across conditions to better understand how these performance differences emerged.

\subsubsection{RQ2: How do users allocate attention and interact with the CUA under different feedback conditions during multitasking?}\label{5_RQ2_results}
\textbf{{\name} enabled more effective intervention when errors occurred, but did not require more attention from users.}

To better understand the performance differences observed in \textbf{RQ1}, we analyzed \textit{spreadsheet error counts} and user interaction behaviors, including \textit{monitoring time} and \textit{task switching}.

A linear mixed-effects model revealed a significant effect of \emph{Condition} on \textit{spreadsheet error counts} (Wald $\chi^2(2)=15.46$, $p<.001$).
Estimated marginal means were 2.51 errors for \textbf{BL}, 2.32 for \textbf{PT}, and 1.31 for \textbf{SK} (Figure~\ref{fig:results}e).
Holm-corrected post-hoc comparisons showed that \textbf{SK} produced significantly fewer errors than both \textbf{BL} ($p<.001$) and \textbf{PT} ($p=.004$), while the difference between \textbf{BL} and \textbf{PT} was not significant.
This suggested that {\name} helped users address CUA errors better than \textbf{BL} and \textbf{PT} in the spreadsheet task.

We then examine whether these improvements resulted from increased monitoring of the CUA by analyzing the \textit{number of task switches} (Figure~\ref{fig:results}g), and the \textit{time spent} monitoring the spreadsheet task (Figure~\ref{fig:results}f).
No significant effects were found for \textit{task switching} (Wald $\chi^2(2)=5.06$, $p=.080$) or \textit{monitoring time} (Wald $\chi^2(2)=2.56$, $p=.278$). Users switched tasks a similar number of times (\textbf{BL}: 13.34, \textbf{PT}: 11.70, \textbf{SK}: 12.48) and spent comparable time monitoring (\textbf{BL}: 12.78s, \textbf{PT}: 10.45s, \textbf{SK}: 8.63s).

These results suggest that \emph{{\name} improved collaboration without increasing monitoring effort}.
Qualitative feedback further supported this finding. 
Participants (N=21) noted that color cues in the \textit{{\name} display} clearly signaled the CUA’s state and enabled timely intervention. As P17 noted, \user{\textbf{SK} was easy ... I didn’t have to focus on both tasks ... I could tell from colors like yellow or red and step in before things got worse.} 
It also reduced the need to review chat logs; as P10 explained, \user{it takes time to go through logs in \textbf{BL}, but color cues help me quickly understand what’s happening and make decisions.}
Other visual cues further improved interpretability. Thumbnails (N=12) and high-level summaries of CUA states (N=19) in \textit{{\name} display} provided contextual grounding, making it \user{easier to tell what the agent is doing than text alone [\textbf{PT}]} (P6) or \user{immediately see how far it has gotten} (P2), while recency indicators (N=13) helped track recent actions without scanning histories, \user{seeing where it was working most recently, so I don’t have to scroll through the chat} (P24).

Together, these feedback explain the improved CUA-assisted task performance in \textbf{RQ1}, enabling users to address errors timely without increased monitoring or extensive browsing of chat logs compared to \textbf{BL} and \textbf{PT}.

\begin{figure*}[h]
\begin{center}
\vspace{-1pc}
\includegraphics[width=0.9\linewidth]{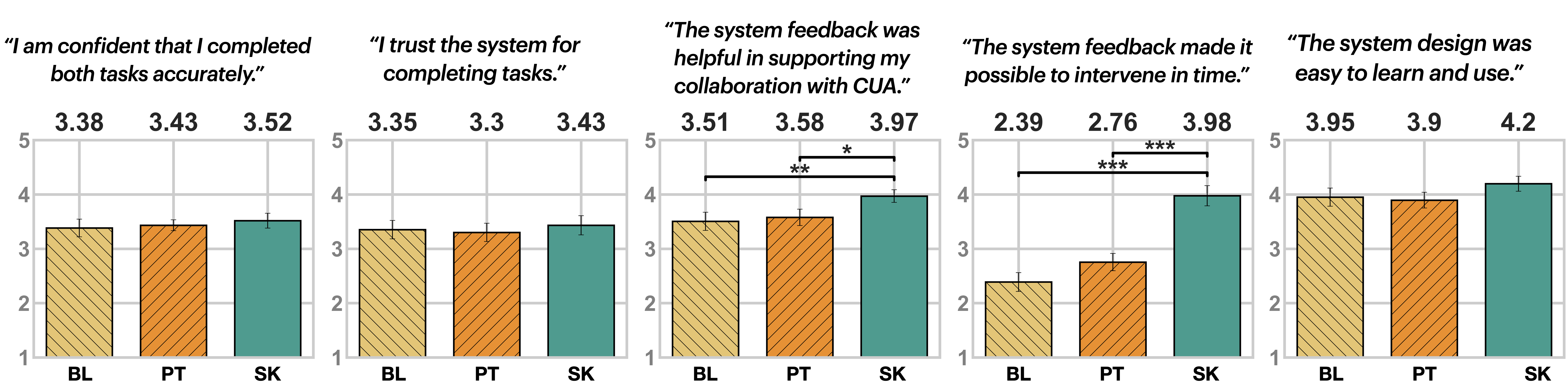}
\vspace{-1pc}
\caption{
Results across conditions (\textbf{BL}, \textbf{PT}, and \textbf{SK}) for five Likert-scale measures (1--5): confidence in completing both tasks accurately, trust in the system for task completion, perceived helpfulness of system feedback for multitasking with the CUA, perceived ability to notice errors in time to intervene, and perceived ease of learning and use. Bars show mean values with error bars indicating $\pm$1 standard error. Statistical significance is indicated as $p<.05$ (*), $p<.01$ (**), and $p<.001$ (***).
}
\label{fig:likert}
\vspace{-1pc}
\Description{Results across conditions (\textbf{MN}, \textbf{BL}, \textbf{PT}, and \textbf{SK}) for five Likert-scale measures (1--5): confidence in completing both tasks accurately, trust in the system for task completion, perceived helpfulness of system feedback for multitasking with the CUA, perceived ability to notice errors in time to intervene, and perceived ease of learning and use. Bars show mean values with error bars indicating $\pm$1 standard error. Statistical significance is indicated as $p<.05$ (*), $p<.01$ (**), and $p<.001$ (***).}
\end{center}
\end{figure*}

\subsubsection{RQ3: How do different feedback conditions influence users’ perceived cognitive load, trust, and overall experience of multitasking with CUAs?}\label{5_RQ3_results}
\textbf{Human-CUA collaboration significantly reduced perceived workload compared to human alone. {\name}'s multimodal feedback was perceived as more helpful for supporting multitasking and enabling timely intervention.}

While \textit{RQ1} and \textit{RQ2} examined objective performance and interaction behaviors, we next assessed participants’ subjective experience using NASA-TLX and post-task Likert ratings.

NASA-TLX results showed that human–CUA collaboration significantly reduced perceived workload compared to working alone (Figure~\ref{fig:nasa}). A linear mixed-effects model revealed a main effect of \emph{Condition} (Wald $\chi^2(3)$, $p<.001$), with higher workload in \textbf{MN} (M=63.83) than in \textbf{BL} (M=52.97), \textbf{PT} (M=56.46), and \textbf{SK} (M=54.11). Post-hoc comparisons confirmed that all CUA-assisted conditions were significantly lower than \textbf{MN}, with no differences among \textbf{BL}, \textbf{PT}, and \textbf{SK}. Similar patterns were observed for \textit{mental demand}, \textit{temporal demand}, and \textit{effort}, while perceived \textit{performance} and \textit{frustration} showed no differences.

Qualitative feedback further supported these findings, where the decreased workload under CUA-assisted conditions was primarily due to task delegation, with users shifting from the role of execution to monitoring. 
For example, P1 noted that the CUA could achieve \user{roughly 80\% accuracy,} transferring their role to \user{verifying revised cells rather than checking every cell entry.}
Similarly, P18 highlighted that the spreadsheet-filling tasks were \user{time-consuming, and offloading them to the agent significantly reduced effort.}
However, monitoring introduced additional effort, as P26 explained, \user{you still have to check the agent... monitoring becomes an extra task}, which may explain the similar workload across CUA-assisted conditions.

Post-task Likert ratings showed similar trust, confidence, and usability across \textbf{BL}, \textbf{PT}, and \textbf{SK} (Figure~\ref{fig:likert}). Perceived trust and confidence were comparable across conditions, likely influenced by the imperfect CUA, which was intentionally designed to introduce errors that required participants to monitor and intervene.
Similarly, no significant differences were found in ease of understanding or use, which is expected given that the feedback mechanisms in \textbf{BL} (chat-only) and \textbf{PT} (peripheral display) were also relatively simple.

However, participants rated \textbf{SK} as significantly more helpful for multitasking than both \textbf{BL} ($p=.005$) and \textbf{PT} ($p=.020$), and as enabling more timely intervention when errors occurred (both $p<.001$).
Qualitative feedback supports these findings, especially for transition support and transparency. 
The \textit{{\name} display} reduced context switching, as P11 noted, \user{it reduced the effort of switching to the background to see how it’s running compared to the \textbf{BL}.}
Upon resuming tasks with the CUA, {\name} helped users quickly understand prior actions; as P22 noted, \user{replay was good for seeing what the agent edited. I usually check the cell it just finished.} Verbal feedback also maintained awareness during transitions, with P23 explaining, \user{when I switched back to the math window, I could still hear what it did, making sure it worked properly after I left.}
Finally, real-time visualization of the CUA’s reasoning improved transparency; as P28 noted, \user{the animation was helpful because I could see what's happening. I feel more confident that the answers are real if I can see how they were derived.}

Together, these results suggested that human–CUA collaboration reduced workload by offloading execution, but still required monitoring, leading to similar workload across CUA conditions. 
Importantly, {\name}’s multimodal feedback did not increase cognitive burden and was perceived as more effective for timely intervention, transparency, and rapid context resumption.

\section{\camera{Proposed} Application Scenarios}
We propose three applications to demonstrate how {\name}'s design principles and interactions could extend beyond spreadsheet tasks to broader workflows. \camera{We acknowledge, however, that their benefits have not yet been systematically evaluated, and we view such evaluation as an important direction for future work.}

\begin{figure}[H]
\centering
\vspace{-0.7pc}
\includegraphics[width=0.9\linewidth]{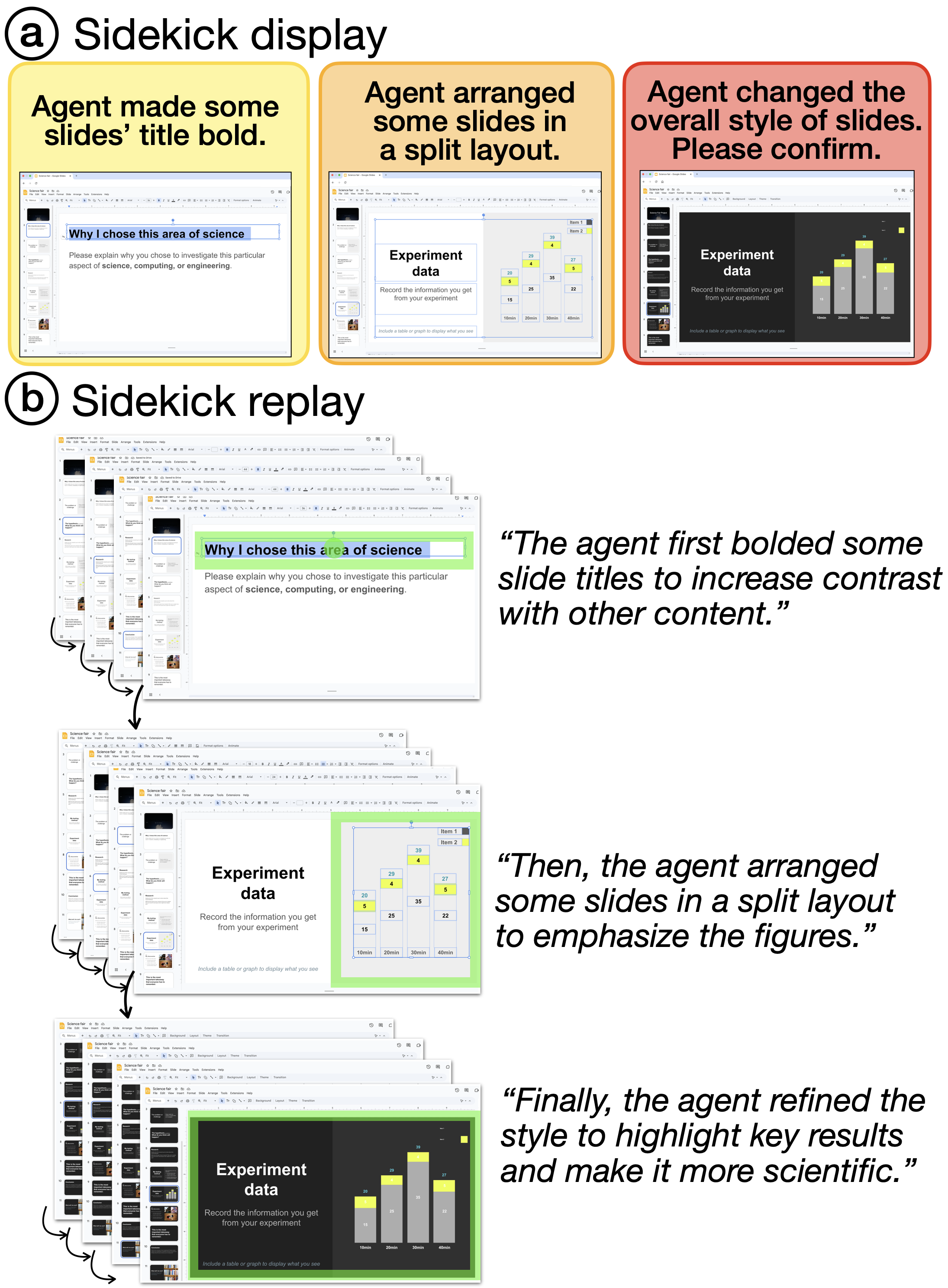}
\vspace{-0.5pc}
\caption{{\name}’s potential application in slide editing.
(a) Color cues in the \textit{{\name} display} can encode the impact of design changes.
(b) {\name} can replay prior visual changes and explain design rationale upon resumption.
}
\label{fig:slideapp}
\Description{The figure is divided into two main sections labeled (a) and (b), illustrating a Sidekick display'' and a Sidekick replay'' of an agent’s actions on presentation slides.
In panel (a), titled Sidekick display,'' three slide-editing actions are shown side by side, each enclosed in a colored rounded rectangle. The first box, with a yellow border, contains the text: Agent made some slides’ title bold.'' Below it is a slide screenshot where the title Why I chose this area of science'' is visibly bolded. The second box, with an orange border, reads: Agent arranged some slides in a split layout.'' The corresponding slide shows a layout with text on one side and a structured diagram or grid on the other, labeled Experiment data.'' The third box, with a red border, reads: Agent changed the overall style of slides. Please confirm.'' The associated slide shows a darker theme with a chart labeled ``Experiment data,'' including vertical bar-like shapes, indicating a visual style change. The progression of colors from yellow to red suggests increasing impact or importance of the agent’s actions.
In panel (b), titled Sidekick replay,'' a sequence of slide screenshots is arranged vertically with slight overlap, representing a replay of the agent’s actions over time. Each step is highlighted with a green rectangular outline around the relevant part of the slide. The first highlighted slide shows the bolded title Why I chose this area of science,'' accompanied by the caption: The agent first bolded some slide titles to increase contrast with other content.'' The second highlighted slide shows the split layout with the Experiment data'' section, accompanied by the caption: Then, the agent arranged some slides in a split layout to emphasize the figures.'' The third highlighted slide shows the final dark-themed design with graphical elements, accompanied by the caption: Finally, the agent refined the style to highlight key results and make it more scientific.'' Arrows between the stacked screenshots indicate the temporal progression of these actions.
}
\end{figure}

\textbf{Supporting complex multimodal tasks (Figure~\ref{fig:slideapp}).}
Future CUAs may enable extended workflows spanning multiple applications. For example, a user might ask a CUA to prepare a presentation by gathering content, inserting charts, and formatting layouts, producing outputs that require timely aesthetic judgment.
In such settings, thumbnails in the \textit{{\name} display} could enhance awareness of images, layouts, and styles, supporting timely intervention and informed decisions. Color cues could extend beyond error signaling to reflect the impact of design changes (Figure~\ref{fig:slideapp}a; \textit{e.g.,} low impact: font changes; high impact: overall style).
Upon return, {\name} could summarize prior actions through verbal explanations and replay, highlighting design rationale, edited elements, and layout changes (Figure~\ref{fig:slideapp}b). 
These capabilities could generalize to other creative multimodal workflows, such as image or video editing.

\begin{figure}[h]
\centering
\includegraphics[width=0.95\linewidth]{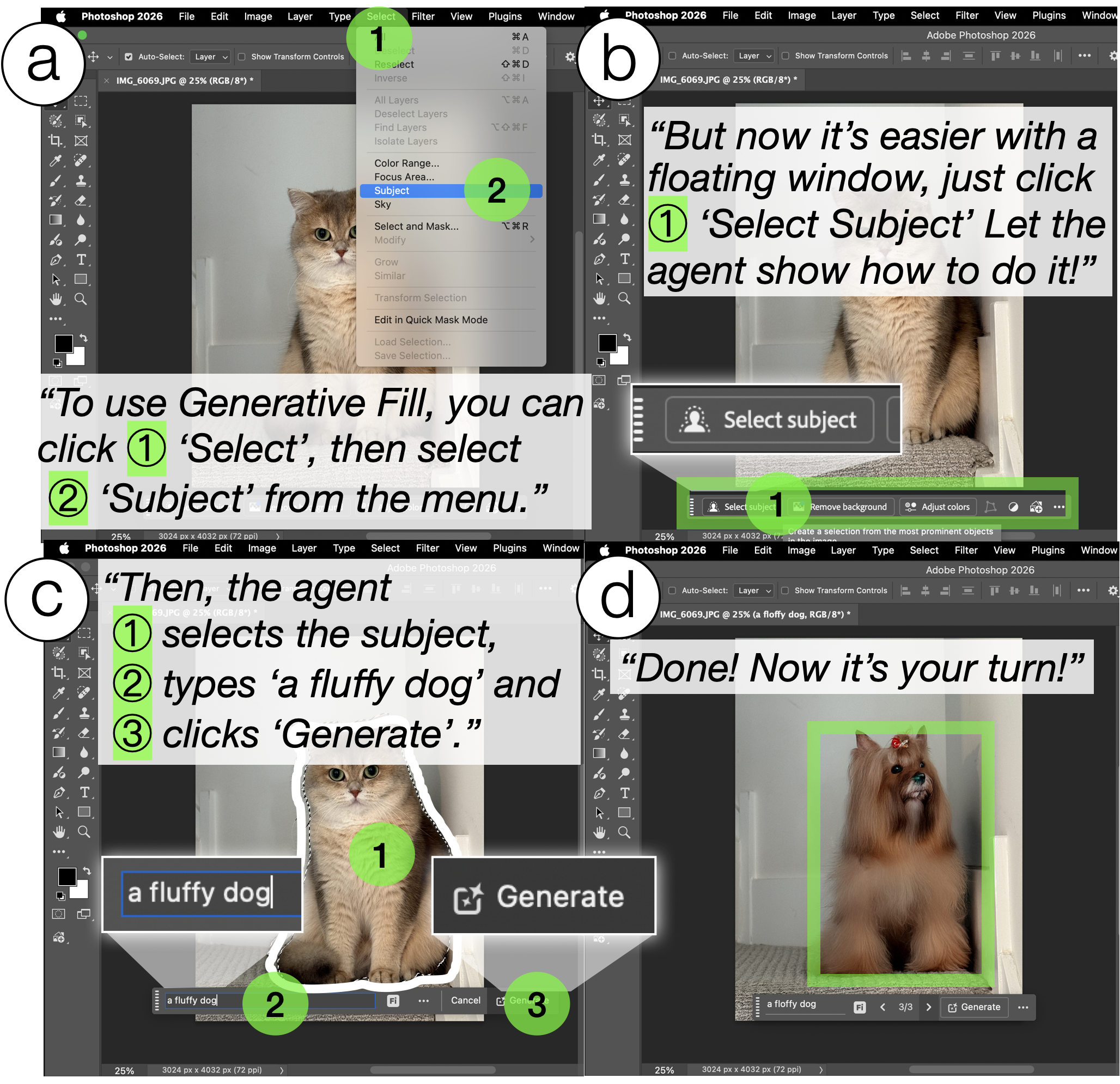}
\vspace{-0.8pc}
\caption{{\name}’s potential application in software tutorials. {\name} could explain how to achieve a function step by step while the CUA demonstrates the actions. Users could observe and learn from the real-time visualizations and verbal feedback.
}
\label{fig:learnapp}
\Description{The figure is divided into three sections labeled (a), (b), and (c), illustrating a human–agent workflow, Sidekick’s display of issues, and how Sidekick reports usability issues through replay.
In panel (a), titled human-agent workflow,'' a diagram shows four main components: a box labeled user,'' a box labeled coding agent,'' and a green rounded rectangle labeled Sidekick,'' inside which there is a smaller yellow box labeled CUA.'' Arrows indicate the flow of interaction. From Sidekick'' to user,'' an arrow labeled report'' points back to the user. From user'' to coding agent,'' a downward arrow is labeled adjust prompt.'' Between coding agent'' and Sidekick,'' there are two arrows: one labeled debrief'' pointing from Sidekick to the coding agent, and another labeled Generative UI'' pointing from the coding agent to Sidekick. Inside the Sidekick box, the CUA'' component has a small loop labeled ``test,'' indicating internal iteration or execution. This panel illustrates how the user, coding agent, and Sidekick interact through reporting, prompt adjustment, and feedback loops.
In panel (b), titled Sidekick display,'' four colored boxes are shown in a row, representing different numbers of detected issues. The boxes are labeled 0 issue'' in green, 3 issues ...'' in yellow, 6 issues ...'' in orange, and ``9 issues ...'' in red. The progression in color from green to red indicates increasing severity or quantity of usability issues identified by Sidekick.
In panel (c), titled Sidekick report usability issues by replay,'' a sequence of interface screenshots is shown on the left, connected by downward arrows to indicate a replay of user or agent interactions. On the right, corresponding textual descriptions of usability issues are provided, each associated with a heuristic label. The first reads: H5 Error prevention: No required field indicators. Agent submitted without entering anything.'' The second reads: H4 Consistency & standards: No links or visual difference. Agent clicked, but nothing shown.'' The third reads: H7 Flexibility & efficiency: No 'view all' option. Required agent to click each tab.'' The fourth reads: H6 Recognition over recall: Initials-overlap. Two 'SC' made agent confused.'' Key phrases such as Agent submitted without entering anything,'' Agent clicked, but nothing shown,'' Required agent to click each tab,'' and ``Two 'SC' made agent confused'' are highlighted in green, emphasizing the specific usability problems encountered. This panel demonstrates how Sidekick uses replay to identify and explain usability issues based on established heuristics.
}
\end{figure}

\textbf{Supporting new interface learning over the CUA's and {\name}'s shoulder (Figure~\ref{fig:learnapp}).}
Learning new software interfaces has long been studied in HCI, with prior work leveraging tutorials from videos and crowd-sourced sources (e.g., blogs)~\cite{Lafreniere2014, lafreniere2013community, GamiCAD, Hudson2018, ambient_help, wang2018leveraging, matejka2009communitycommands}. 
As AI increasingly enables users to generate applications and interfaces in natural language, learning new software or interfaces developed by others is likely to become more common.
{\name} offers a learning opportunity when integrated with CUAs: by providing real-time visualizations and verbal feedback of CUA reasoning, it enables users to observe and learn from the agent’s actions. For example, when prompted with a tutorial on features like “Generative Fill,” {\name} can explain steps sourced from the online community while the CUA demonstrates them in real time (Figure~\ref{fig:learnapp}).

\begin{figure}[h]
\centering
\includegraphics[width=0.9\linewidth]{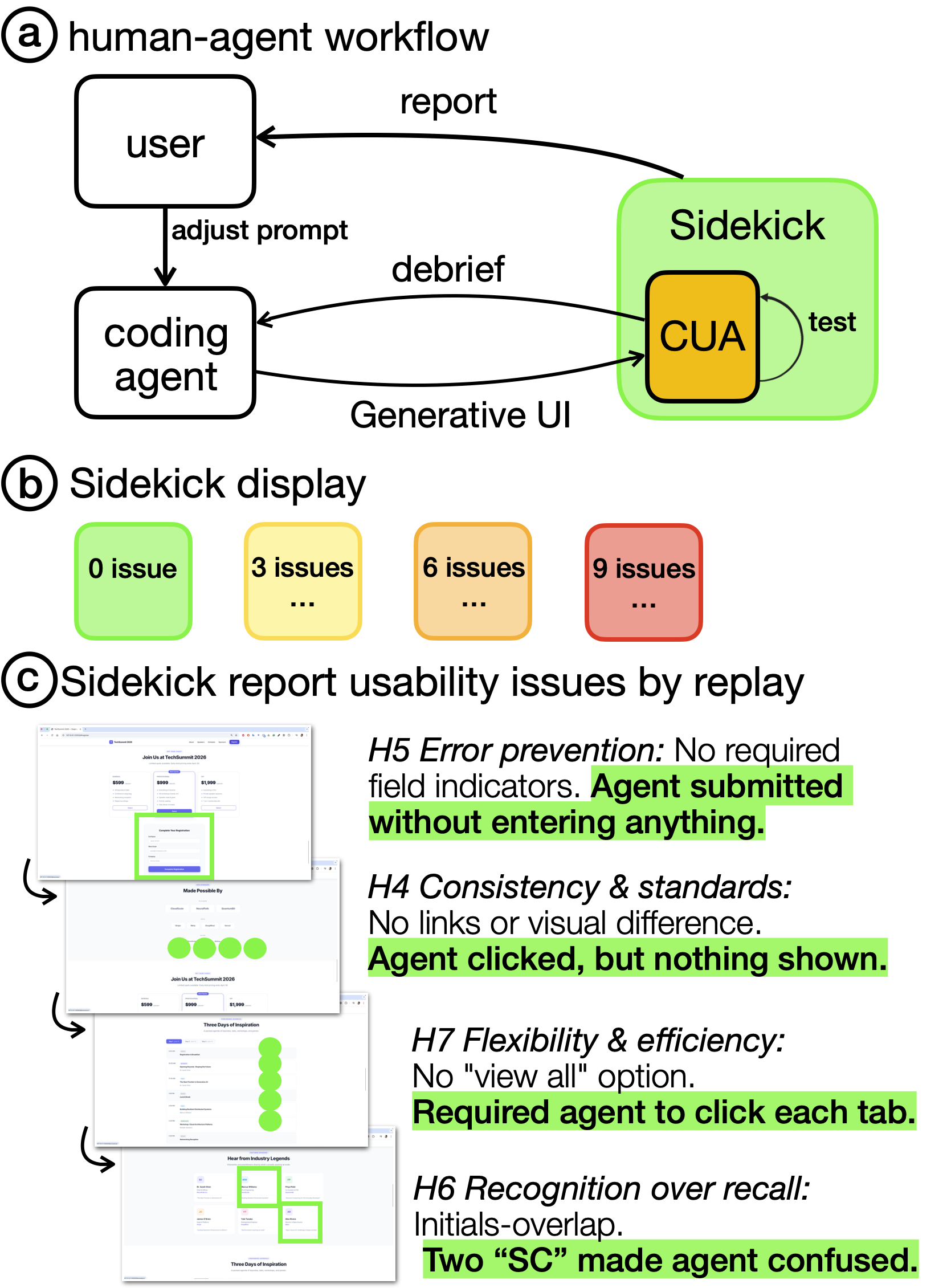}
\vspace{-0.5pc}
\caption{{\name}’s potential application in collaboration with coding agents and CUAs for a usability test report.
(a) Integration into closed-loop workflows for UI generation and evaluation.
(b) Color cues in the \textit{{\name} display} encode the number or severity of identified usability issues.
(c) {\name} reports key results by explaining issues alongside relevant guidelines.
}
\label{fig:uiapp}
\Description{The figure consists of four panels labeled (a), (b), (c), and (d), each showing a step in using a generative editing feature within an image editing software interface (Adobe Photoshop 2026). The panels depict a progression of actions performed with assistance from an agent.
In panel (a), a screenshot of the Photoshop interface shows an image of a cat centered on the canvas. At the top menu bar, a green circular marker labeled 1'' highlights the menu item Select.'' A dropdown menu is open, and another green marker labeled 2'' highlights the option Subject.'' Below the image, a caption reads: ``To use Generative Fill, you can click (1) ‘Select’, then select (2) ‘Subject’ from the menu.'' This panel explains the initial step of selecting the subject in the image.
In panel (b), the Photoshop interface is shown again with the same cat image. A text overlay reads: But now it’s easier with a floating window, just click (1) ‘Select Subject’ Let the agent show how to do it!'' A floating button labeled Select subject'' appears near the image, with a green marker ``1'' indicating where to click. This panel introduces a simplified interaction using a floating control.
In panel (c), the interface shows the cat image with a selection outline around it. A text overlay states: Then, the agent (1) selects the subject, (2) types ‘a fluffy dog’ and (3) clicks ‘Generate’,'' with green markers labeled 1, 2, and 3 corresponding to each step. At the bottom of the interface, a text input field contains the phrase a fluffy dog,'' and a button labeled ``Generate'' is visible next to it. This panel demonstrates the agent performing a generative edit by specifying a prompt and executing the action.
In panel (d), the final result is shown: the cat image has been replaced with an image of a fluffy dog. A green rectangular highlight surrounds the generated dog image. A caption at the top reads: ``Done! Now it’s your turn!'' This panel presents the outcome of the generative fill process.
}
\end{figure}

\textbf{Reporting usability tests of generative UIs to users (Figure~\ref{fig:uiapp}).}
GUI testing is essential for usability and accessibility~\cite{nielsen1990heuristic, AXNav, Swearngin2024}. Agentic platforms can support closed-loop workflows where coding agents generate GUIs and CUAs evaluate them against heuristic or accessibility guidelines~\cite{claudeCUAcowork}. However, communication between agents and users remains limited.
{\name} could bridge this gap by surfacing usability issues in real time through its display or providing detailed multimodal reports upon user engagement (Figure~\ref{fig:uiapp}a).
It may also track recurring errors or usability issues and summarize them for users after several iterations.
Users can intervene by monitoring the \textit{{\name} display}, where color cues reflect issue severity (Figure~\ref{fig:uiapp}b), signaling when prompt revisions are needed. 
Additionally, summaries upon return, along with real-time visualization and verbal feedback, can help users verify testing rationale against intended guidelines (Figure~\ref{fig:uiapp}c), improving transparency and confidence.

\section{Discussion and Future Work}
We discuss the limitations, our lessons learned, and implications for extending {\name} in broader scenarios and future CUAs.

\textbf{Design implications for multimodal human-CUA communication.}
In our study, {\name} improved multitasking with CUAs over baseline systems via multimodal feedback to communicate agent states. 
Recent work by Cheng et al.~\cite{Cheng_2026} proposed a taxonomy of human-CUA interactions; {\name} partially aligns with this taxonomy by making agent activities visible, increasing transparency of reasoning, communicating runtime status, and enabling user intervention when errors occur.
However, feedback design still requires additional user-centered refinement. 
For example, preferences for feedback modality varied: some participants found visual feedback (e.g., color changes) helpful but distracting (Section~\ref{5_RQ1_results}), and suggested lighter alternatives such as brief audio cues. In contrast, in foreground interactions, some noted that visual feedback alone may suffice, while speech was unnecessary.
These findings highlighted the need for customizable feedback and context awareness, which could draw on prior design principles or systems in attention management, such as tuning information density to manageable chunks~\cite{sawhney2000nomadic, ROPE, dabbish_kraut2004, WorldScribe, TouchScribe}, adapting notification modality to user context~\cite{somervell2002evaluating, Maglio2000, Brewster1994, SoundShift, arroyo2002interruptions, Viago}, and delivering information at opportune moments~\cite{Chen2026Opportune, Oasis, Iqbal2008, BusyBody, Lilsys}.
Finally, beyond conveying CUAs' operation states, future work could further facilitate bidirectional human-CUA alignment by communicating user contexts to CUAs (e.g., goals, safety, and privacy constraints) and clarifying CUAs' capabilities and knowledge scope to users~\cite{Amershi2019, Cheng_2026}. 

\textbf{Supporting human-agent communication at scale.}
In this work, we developed {\name} to address the communication gap between humans and CUAs during multitasking, with a focus on enhancing error awareness in our study. 
\camera{Rather than surfacing every error, {\name} acts as a monitoring proxy that requests user intervention only when accumulated errors exceed a threshold (Section~\ref{4_system_background}). This enables strategic check-ins while allowing CUAs to recover independently from minor issues.}

However, as CUAs become faster and more accurate, users’ communication needs may shift beyond error detection toward subjective concerns such as aesthetics, preferences, and privacy. In such cases, {\name} could therefore support richer interactions that communicate nuanced decisions and outcomes, as our application scenarios demonstrated (Figure~\ref{fig:slideapp}).
As CUA speed increases, users may also face a surge of outputs in a short time. To address this, {\name} could provide hierarchical summaries that first surface high-stakes artifacts  (e.g., key slides among many outputs) or information likely to contain errors or expectation mismatches, and allow users to drill down into other details if needed.

Moreover, {\name} currently focuses on a single CUA performing one task within one window, whereas real-world settings may involve multiple agents operating across different workspaces as such systems become more capable and widespread. This direction aligns with prior work on unified interfaces for managing multiple workspaces~\cite{jiang2025orca, CodeBubbles, ElasticWindows, tabsdo}.
Accordingly, multi-workspace environments may become the primary context for {\name}, supporting coordination across multiple agents. The {\name} display could shift from per-agent views to aggregated summaries (e.g., overall progress indicators or selective visual snippets highlighting issues~\cite{WinCuts}), with color cues encoding cross-agent states such as conflicts or dependencies.
Upon returning, {\name} could provide selective summaries of cross-agent activities, prioritizing information based on user goals. In foreground scenarios, it could highlight critical agents or visualize dependencies to help users detect conflicts and coordinate interventions.

\camera{
Beyond supporting users during task execution, recent agent trajectory-analysis systems have transformed lengthy execution traces into behavioral categories, diagnostic measures, and visual representations for post-hoc inspection~\cite{gao2026interpretagentbehavior, docent, trace, agentdiagnose}. Building on these approaches, {\name} could be extended to translate low-level human-CUA activities into structured reports that surface consequential behaviors, recurring risks and errors, the effects of user interventions, and moments when additional intervention may be warranted. Such reports could help identify systematic failure patterns and inform improvements to monitoring and intervention strategies. 
More broadly, in the future, {\name} could evolve into a scalable communication layer for increasingly complex and capable multi-agent workflows, providing real-time feedback tailored to users’ goals, attention, and workload, alongside post-hoc reports that support retrospective analysis, auditing, and iterative system improvement. 
}

\textbf{Study limitations.}
As noted in Section~\ref{3_study}, CUAs at the time of our study had limited reliability, leading us to use more controlled tasks (e.g., spreadsheet filling). These tasks reflect common use cases observed in commercial demonstrations~\cite{claudeCUAdemo} and our formative study, which \textit{users can perform the work themselves but prefer to delegate it due to its tedious and repetitive nature.}
While CUA capabilities are expected to improve, our findings may still generalize to similar repetitive tasks (e.g., tax filing or slide formatting), as noted by participants.
Additionally, although reward and penalty mechanisms were calibrated through pilot studies, they may have influenced participants’ mental models and strategies. 
However, we did not observe strong evidence of altered interaction patterns that impact study results (e.g., abandoning CUA outputs or ignoring errors). 
Future work could examine tasks with varying risk levels or alternative incentive structures to better understand their effects on user behavior, trust, and confidence.

\section{Conclusion}
In this work, we introduce {\name}, a prototype system that supports human-CUA communication during multitasking. {\name} provides peripheral signals for background awareness, structured multimodal summaries for context resumption, and real-time synchronized feedback for foreground transparency.
Through a mixed-methods study with 30 participants, we show that {\name} significantly improves multitasking performance over working alone and text-based baselines by enhancing awareness of CUA progress and enabling timely intervention without constant monitoring or disrupting primary tasks.
We further demonstrate its potential in broader scenarios, including multimodal task support, reporting usability issues in multi-agent workflows, and enabling learning through agent demonstrations. Finally, we discuss implications for more customizable, lightweight human–CUA communication and the scalability of {\name} as CUAs continue to improve.

\begin{acks}
We thank all study participants and anonymous reviewers for their valuable feedback, as well as Prof. Xu Wang for suggestions on the experimental design and analysis. This research was supported in part by an Adobe Research Gift and a Google Academic Research Award. Ruei-Che Chang was also supported by the Apple Scholars in AI/ML PhD Fellowship.
\end{acks}

\bibliographystyle{ACM-Reference-Format}
\bibliography{main}

\appendix
\onecolumn

\newpage
\begin{table}
\caption{Participants in our study were marked as P1-P30.}
\centering
\vspace{-.8pc}
\label{tab:demographic}
\begin{tabular}{|l|l|l|l||l|l|l|l|}
    \hline
    \textbf{ID} & \textbf{Age} & \textbf{Gender} & \textbf{Major/Profession} & \textbf{ID} & \textbf{Age} & \textbf{Gender} & \textbf{Major/Profession}\\
    \hline
    P1  & 24 & Female & Data Science                             & P16 & 24 & Female & Computer Science                     \\
    \hline
    P2  & 25 & Female & Public Health                            & P17 & 26 & Female & Electrical  Computer Engineering     \\
    \hline
    P3  & 23 & Female & Quantitative Finance  Risk Management    & P18 & 28 & Male   & Computer Science and Engineering     \\
    \hline
    P4  & 22 & Male   & Computer Science and Engineering         & P19 & 22 & Female & Computer Science and Engineering     \\
    \hline
    P5  & 25 & Female & Nutritional Science                      & P20 & 19 & Female & Electrical and Computer Engineering  \\
    \hline
    P6  & 23 & Male   & environment and sustainability           & P21 & 22 & Male   & Computer Science and Engineering     \\
    \hline
    P7  & 26 & Male   & Solution Intellgence Analyst             & P22 & 26 & Male   & Electrical and Computer Engineering  \\
    \hline
    P8  & 32 & Male   & Postdoc                                  & P23 & 19 & Male   & Computer Science and Engineering     \\
    \hline
    P9  & 24 & Female & Geospatial data science                  & P24 & 24 & Male   & Computer Science and Engineering     \\
    \hline
    P10 & 23 & Female & Quantitative Finance and Risk Management & P25 & 26 & Female & Industrial Engineering               \\
    \hline
    P11 & 23 & Male   & Computer Science and Engineering         & P26 & 25 & Male   & Electrical and Computer Engineering  \\
    \hline
    P12 & 25 & Female & Computer Science and Engineering         & P27 & 23 & Male   & Computer Science and Engineering     \\
    \hline
    P13 & 20 & Female & Anthropology and astronomy               & P28 & 18 & Male   & Computer Science and Engineering     \\
    \hline
    P14 & 22 & Female & Computer Science and Engineering         & P29 & 23 & Female & Computer Science and Engineering     \\
    \hline
    P15 & 28 & Female & Information Science                      & P30 & 23 & Female & Accountant     \\ 
    \hline
    \end{tabular}
\end{table}

\end{document}